\documentclass[aps,pra,twocolumn,superscriptaddress,groupedaddress]{revtex4-2}
\usepackage[colorlinks=true,citecolor=blue,linkcolor=red,urlcolor=red]{hyperref}
\usepackage[sort&compress]{natbib}
\usepackage{graphicx}
\usepackage{amssymb}
\usepackage{braket}
\usepackage{amsmath}
\usepackage{hyperref}
\usepackage{dsfont}
\usepackage{bm} 
\usepackage{physics}
\usepackage{tabularx}
\usepackage{verbatim}
\usepackage{float}
\usepackage{romannum}

\renewcommand\k{{\bf k}}
\renewcommand\r{{\bf r}}
\newcommand\p{{\bf p}}

\newcommand\x{{\bf x}}

\begin{document}
	
\author{Anirudh Gundhi}
\email{anirudh.gundhi@phd.units.it}
\affiliation{Department of Physics, University of Trieste, Strada Costiera 11, 34151 Trieste, Italy}
\affiliation{Istituto Nazionale di Fisica Nucleare, Trieste Section, Via Valerio 2, 34127 Trieste, Italy}
		
\author{Angelo Bassi}
\email{abassi@units.it}
\affiliation{Department of Physics, University of Trieste, Strada Costiera 11, 34151 Trieste, Italy}
\affiliation{Istituto Nazionale di Fisica Nucleare, Trieste Section, Via Valerio 2, 34127 Trieste, Italy}

\title{Motion of an electron through vacuum fluctuations}
	
\date{\today}

\begin{abstract}
We study the effects of the electromagnetic vacuum on the motion of a nonrelativistic electron. First, we derive the equation of motion for the expectation value of the electron's position operator. We show how this equation has the same form as the classical Abraham-Lorentz equation but, at the same time, is free of the well known runaway solution. Second, we study decoherence induced by vacuum fluctuations. We show that decoherence due to vacuum fluctuations that appears at the level of the reduced density matrix of the electron, obtained after tracing over the radiation field, does not correspond to actual irreversible loss of coherence.
\end{abstract}

\maketitle
\section{Introduction}
Numerous physical phenomena such as the Casimir effect \cite{Casimir1948, Birrell, parker_toms_2009}, the Unruh effect \cite{UnruhEffect, Fulling, Takagi} and the Lamb shift \cite{BetheLambShift, LambLambShift, Welton, dalibard} are attributed to the presence of vacuum fluctuations. The possibility of decoherence due to vacuum fluctuations, as being fundamental and unavoidable, has also been discussed in various works \cite{KieferDecoherence, Kiefer:1992, Ford, Baym_Ozawa, Santos1994, Diosi1995, Caldeira1991, BandP} without arriving at a general consensus.

Similarly, the quantum mechanical version of the classical Abraham-Lorentz (AL) equation, which describes the recoil force experienced by an accelerated electron due to the emission of radiation \cite{Coleman1982,Pearle1982,Griffiths2017,TannoudjiPartOneChapterOne}, has been previously derived, for example, in \cite{dalibard}.  However, the equation was obtained for the position operator of the electron and its direct connection with the classical dynamics was found to be difficult to make. This was due to the presence of the additional transverse electric field operator of the electromagnetic vacuum, which is zero classically. A similar problem persists concerning the interpretation of the quantum Langevin equation obtained in \cite{Caldeira1991} for an electron interacting with vacuum fluctuations. We also refer to \cite{Hu2002} and references therein for related works within the context of scalar quantum electrodynamics.

In this article, working within the framework of open quantum systems, treating the electromagnetic (EM) field as the environment and the electron as the system of interest, we obtain the expression for the time evolution of the reduced density matrix of the electron, in the position basis, after tracing over the environment. The formalism used is adopted from \cite{calzetta_hu_2008}. Using the master equation, we obtain the equation of motion (EOM) for the expectation value of the position operator which, in contrast to \cite{dalibard,Caldeira1991}, provides a direct correspondence with the classical dynamics. In the presence of an arbitrary potential, we show that the reduced quantum dynamics is the same as the classical AL equation. Moreover, the equation that emerges after a quantum mechanical treatment appears to be free of the problems associated with the AL equation: the runaway solution which leads to an exponential increase of the electron's acceleration, even in the absence of an external potential \cite{Coleman1982,Pearle1982,Griffiths2017}. 

Further, we show that the loss of coherence due to vacuum fluctuations at the level of the reduced density matrix is only apparent and reversible. To this end, we show that by ``switching off'' the interactions with the EM field, the original coherence is restored at the level of the system. Moreover, the expression for the decoherence factor that we obtain differs from the ones obtained in \cite{Caldeira1991,BandP}, where the authors argue for a finite loss of coherence for momentum superpositions, due to vacuum fluctuations, but with different estimates for the magnitude of decoherence.

Here, we outline the main results of the article. In Sec.~\ref{Sec:Dynamics}, starting from the Lagrangian of the nonrelativistic electron [cf.  Eq.~\eqref{eq:Lag}], we obtain the corresponding Hamiltonian. The main result of this section is the identification of the full effective potential governing the dynamics [cf. Eqs.~\eqref{eq:Vfull} and~\eqref{eq:VEMExact}]. Then, upon standard quantization, in Sec.~\ref{Sec:MasterEquation}  we obtain the master equation for the effective reduced density matrix of the nonrelativistic electron up to second order in the interactions [cf. Eq.~\eqref{eq:MasterEquation}]. This is the main technical result of the article. The noise and the dissipation kernels, which appear in this equation, are also derived explicitly in Sec.~\ref{Sec:Kernels} [cf. Eqs.~\eqref{eq:NoiseKernel} and~\eqref{eq:DissipationKernel}]. Next, with the help of the master equation, in Sec.~\ref{Sec:EOM} we obtain the equation of motion for the expectation value of the electron's position operator [cf. Eqs.~\eqref{eq:AvX} and~\eqref{eq:AvPTrace}]. We show  how the EOM that we derive, after a full quantum treatment, is free of the well-known problems related to the classical AL equation. This is one of the two main physical results of the article.  Finally, in Sec.~\ref{Sec:Decoherence}, we discuss decoherence due to vacuum fluctuations. Although, on the face of it, the master equation suggests finite decoherence due to vacuum fluctuations [cf. Eqs.~\eqref{eq:Decoherence} and~\eqref{eq:Coherence_Length}], we show that this decoherence effect is only apparent and that the electron never looses coherence irreversibly to its environment [cf. Eq.~\eqref{eq:falsedec}]. This is the other main physical result of the article.
\section{The Lagrangian and the Hamiltonian formalism} \label{Sec:Dynamics}
We begin  by formulating the Lagrangian and the Hamiltonian relevant for the dynamics of a nonrelativistic electron in the presence of an external potential and an external radiation field.
\subsection{The Lagrangian}\label{Sec:Lagrangian}
In the Coulomb gauge, the standard Lagrangian for the dynamics of a nonrelativistic electron in the presence of an external potential and an external radiation field is given by \cite{TannoudjiPartOneChapterTwo}
\begin{align}\label{eq:TotLagCoulomb}
L ={}& \frac{1}{2}m\dot{\r}_{e}^2 -V_0(\r_e, t)+\frac{\epsilon_{0}}{2}\int d^3r\left(\mathbf{E}_{\perp}^2(\r,t)-c^2\mathbf{B}^2(\r,t)\right)\nonumber\\&+\int d^3r\mathbf{j}(\r,t)\cdot\mathbf{A}_{\perp}(\r,t)- \int_{1/2} d^3k\frac{|\rho|^2}{\epsilon_{0}k^2}\,.
\end{align}
Here, $\textbf{r}_e$ denotes the position of the electron, $m$  the bare mass, $e$ the electric charge,  $V_0$ an arbitrary bare external potential acting only on the electron, $\mathbf{E}_{\perp}$ the transverse electric field (obtained by taking the negative partial time derivative of the vector potential $\mathbf{A}_{\perp}$), $\mathbf{B}$ the magnetic field (obtained by taking the curl of $\mathbf{A}_{\perp}$), $\epsilon_{0}$ the permittivity of free space, $c$ the speed of light, $\rho$ the charge density, and $\mathbf{j}$ the corresponding current density. 

The last term in Eq.~\eqref{eq:TotLagCoulomb} describes the Coulomb potential between different particles which is written in Fourier space, where the symbol $\int_{1/2}$ means that the integral is taken over half the volume in the reciprocal space. For a single particle, it reduces to the particle's Coulomb self-energy $E_{\mathrm{Coul}}$. After the introduction of a suitable UV cutoff ($\omega_{\text{\tiny{max}}}$),  which is also necessary for the calculations that are to follow (cf. Secs.~\ref{Sec:Hamiltonian} and~\ref{Sec:Kernels}), it takes a finite value given by $E_{\mathrm{Coul}}=\alpha\hbar\omega_{\text{\tiny{max}}}/\pi$ (cf. Eq.~(B.36)  in \cite{TannoudjiPartOneChapterOne}). This term is ignored in our analysis since it is a constant and does not affect the motion of the electron.

For the electron, the current density is given by $\mathbf{j}(\r,t) = -e\dot{\r}\delta(\r-\r_{e})$ and therefore, the interaction term in the second line of Eq.~\eqref{eq:TotLagCoulomb} becomes $-e\dot{\r}_e\mathbf{A}_{\perp}(\r_e,t)$. After integrating by parts inside the action, this term can be written as
\begin{align}\label{eq:IntegrationByParts}
-e\int dt \dot{\r}_e\mathbf{A}_{\perp} ={} &e\int dt \left(\r_e\cdot \frac{d}{dt} \mathbf{A}_{\perp}-\frac{d}{dt}(\r_{e}\cdot \mathbf{A}_{\perp})\right)\,.
\end{align}
The total derivative $d/dt(\r_{e}\cdot\mathbf{A}_{\perp})$ inside the Lagrangian does not affect the dynamics and is therefore ignored in our analysis (see, also, \cite{Caldeira1991}). Further, the total time derivative acting on $\mathbf{A}_{\perp}(\r_{e},t)$ in the first term on the right-hand side of Eq.~\eqref{eq:IntegrationByParts} can be replaced with a partial time derivative. This is because, even though
\begin{align}\label{eq:TotalDerivative}
\frac{d}{dt}\textbf{A}_{\perp}(\r_e(t),t)=\partial_t\textbf{A}_{\perp}+ v^i\partial_i\textbf{A}_{\perp}\,,\qquad v^i:=\dot{\mathrm{r}}_e^i\,,
\end{align}
from the plane-wave solution of $\textbf{A}_{\perp}(\r_e,t)$, the term  $v^i\partial_i\textbf{A}_{\perp}(\r_e,t)$ is seen to be negligible with respect to $\partial_t{{\mathbf{A}}}_{\perp}(\r_{e},t)=-\mathbf{E}_{\perp}(\r_{e},t)$ as long as $\omega_k\gg v k$ or, equivalently, $v\ll c$. Therefore, for an electron traveling at speeds $v\ll c$, the Lagrangian relevant for the dynamics reduces to
\begin{align}\label{eq:Lag}
L \approx {}&\frac{1}{2}m\dot{\r}_e^2-V_0(\r_e,t)+\frac{\epsilon_{0}}{2}\int d^3r\left(\mathbf{E}_{\perp}^2-c^2\mathbf{B}^2\right)\nonumber\\
&-e\r_{e} \mathbf{E}_{\perp}(\r_{e},t)\,.
\end{align}    
In Eq.~\eqref{eq:Lag}, as mentioned before,  the total derivative $d/dt(\r_{e}\mathbf{A}_{\perp})$ and the constant Coulomb self energy term have been omitted as these do not affect the electron's dynamics.

\subsection{The Hamiltonian}\label{Sec:Hamiltonian}
In terms of the canonical variables $\r_e,\p,\mathbf{A}_{\perp}$ and $\mathbf{\Pi}$, with $\p$ and $\mathbf{\Pi}$ being the conjugate momentums for the variables $\r_e$ and $\mathbf{A}_{\perp}$ respectively, the Hamiltonian corresponding to the Lagrangian~\eqref{eq:Lag} can be written in the form
\begin{align}\label{eq:Hfull}
\text{H}:= \text{H}_{\text{\tiny{S}}}+\text{H}_{\text{\tiny{EM}}}+\text{H}_{\text{int}}\,.
\end{align}
To write its explicit expression, we first define the quantity $\mathbf{\Pi}_{\text{\tiny{E}}}:=-\mathbf{\Pi}/\epsilon_0$ since it appears repeatedly in the calculations. The different components of the full Hamiltonian can then be written as 
\begin{align}
\text{H}_{\text{\tiny{EM}}}= \int d^3r\mathcal{H}_{\text{\tiny{EM}}}=\frac{\epsilon_{0}}{2}\int d^3r(\mathbf{\Pi}_{\text{\tiny{E}}}^2+c^2\mathbf{B}^2)\,,
\end{align}
which is the free-field Hamiltonian of the radiation field, 
\begin{align}
\text{H}_{\text{int}}=e\r _e\mathbf{\Pi}_{\text{\tiny{E}}}(\r _e,t)\,,
\end{align}
which is the term that encodes the interaction between the electron and the radiation field and 
\begin{align}\label{eq:Hs}
\text{H}_{\text{\tiny{S}}}={}&\frac{\p^2}{2m}+{V}_0(\r_e,t)\nonumber\\
&+\frac{e^2}{2\epsilon_{0}}\int d^3rr^i \delta^{\perp}_{im}(\r-\r_e)\delta^{\perp}_{mj}(\r-\r_e)r^j\,,
\end{align}
which is the ``system" Hamiltonian that contains only the canonical variables of the electron.  The transverse Dirac delta  $\delta^{\perp}_{ij}(\r-\r_e)$ that appears in the expression of $\text{H}_{\text{\tiny{S}}}$ is defined to be \cite{TannoudjiPartOneChapterOne}
\begin{align}\label{eq:TD}
\delta^{\perp}_{ij}(\r-\r_e) := \frac{1}{(2\pi)^3}\int d^3k \left(\delta_{ij}-\frac{k_ik_j}{k^2}\right)e^{i\k\cdot(\r-\r_e)}\,.
\end{align}
It appears instead of the Dirac delta due to the coupling of the position of the electron with the transverse electric field in Eq.~\eqref{eq:Lag}. The form of $\text{H}_{\text{\tiny{S}}}$ calls for an identification of the full effective potential $V(\r_e,t)$ governing the dynamics of the electron such that
\begin{align}\label{eq:Vfull}
V(\r_e,t):={}&V_0(\r_e,t)+V_{\text{\tiny{EM}}}(\r_e)\,,\nonumber\\ 
V_{\text{\tiny{EM}}}(\r_e):={}& \frac{e^2}{2\epsilon_{0}}\int d^3rr^i \delta^{\perp}_{im}(\r-\r_e)\delta^{\perp}_{mj}(\r-\r_e)r^j\,.
\end{align}
It should be emphasized that the extra term $V_{\text{\tiny{EM}}}(\r_e)$ is not added to the bare potential by hand, but arises due to the $\r_e\mathbf{E}_{\perp}$ coupling  in the Lagrangian~\eqref{eq:Lag}. Although it gives a divergent contribution $\frac{e^2}{2\epsilon_0}\delta^{\perp}_{ij}(\mathbf{0})r^i_er^j_e$, after regularizing the transverse Dirac delta on a minimum length scale $r_{\mathrm{min}}=1/k_{\mathrm{max}}$, the contribution coming from this term becomes finite.  To be more precise, we impose the UV cutoff consistently in our calculations (cf. Sec.~\ref{Sec:Kernels}) by introducing the convergence factor $e^{-k/k_{\text{\tiny{max}}}}$ inside the Fourier space integrals . Using this procedure, the expression for $\delta^{\perp}_{ij}(\mathbf{0})$ is obtained to be
\begin{align}
\delta^{\perp}_{ij}(\mathbf{0})= \frac{1}{(2\pi)^3}\int dk k^2e^{-k/k_{\text{\tiny{max}}}}\int d\Omega \left(\delta_{ij}-\frac{k_ik_j}{k^2}\right)\,.
\end{align}
First evaluating the angular integral, which gives a factor $\frac{8\pi}{3}\delta_{ij}$, and then the radial integral, we get
\begin{align}\label{eq:VEMExact}
V_{\text{\tiny{EM}}}(\r_e)= \frac{e^2\omega^3_{\text{\tiny{max}}}}{3\pi^2\epsilon_0c^3}\r^2_e\,,
\end{align}
where $\omega_{\text{\tiny{max}}}= ck_{\text{\tiny{max}}}$. As shown in  Sec.~\ref{Sec:EOM}, the potential $V_{\text{\tiny{EM}}}(\r_e)$ plays an important role as it cancels the  contribution coming from another term (up to second order in the interactions), which appears later in the calculations, yielding a consistent EOM for the nonrelativistic electron.

\section{The master equation}\label{Sec:MasterEquation}
The probability amplitude for a particle to be at the position $x_{\text{\tiny{f}}}$ at some final time $t$, starting from the position $x_i$ at some initial time $t_i$, is given by \cite{altland_simons_2010}

\begin{align}\label{eq:TimeDynamics}
\bra{x_{\text{\tiny{f}}}}\hat{U}(t;t_i)\ket{x_i} =&{} \int_{\substack{{x(t) = x_{\text{\tiny{f}}},}\\ {x(t_i) = x_i}}}D[x,p]e^{-\frac{i}{\hbar}\int_{t_i}^{t} dt'\left( \text{H}[x,p]-p\dot{x}\right)}\nonumber\\=&{}\int_{\substack{{x(t) = x_{\text{\tiny{f}}},}\\ {x(t_i) = x_i}}}D[x]e^{\frac{i}{\hbar} S[x]}\,,
\end{align}
where $\text{H}$ is the full Hamiltonian and $S$ is the corresponding action describing some general dynamics. From Eq.~\eqref{eq:TimeDynamics}, the expression for the density matrix at time $t$ can be written as \cite{calzetta_hu_2008}
\begin{align}\label{eq:DensityMatrixGeneral}
\bra{x^{\prime}_{\text{\tiny{f}}}}\hat{\rho}(t)\ket{x_{\text{\tiny{f}}}}  ={}& \int_{\substack{{x(t) = x_{\text{\tiny{f}}} ,}\\ {x^{\prime}(t)= x^{\prime}_{\text{\tiny{f}}}}}}D[x,x^{\prime}]e^{\frac{i}{\hbar}(S[x^{\prime}]-S[x])}\rho^i\,,
\end{align}
where $\rho^i:=\rho(x'_i,x_i,t_i)$ and the integrals over $x_i$ and $x'_i$ are included within the path integral. The expression analogous to Eq.~\eqref{eq:TimeDynamics} also exists for $\bra{p_{\text{\tiny{f}}}}\hat{U}(t;t_i)\ket{p_i}$ in which the boundary conditions are fixed on $p(t)$ and the phase-space weighing function is instead given by $\exp\lbrace\frac{-i}{\hbar}\int_{t_i}^{t}dt'\left(\text{H}[x,p]+x\dot{p}\right)\rbrace$ such that
\begin{align}\label{eq:TransitionMomentum}
\bra{p_{\text{\tiny{f}}}}\hat{U}(t;t_i)\ket{p_i} = \int_{\substack{{p(t) = p_{\text{\tiny{f}}},}\\ {p(t_i) = p_i}}}D[x,p]e^{-\frac{i}{\hbar}\int_{t_i}^{t} dt'\left( \text{H}[x,p]+x\dot{p}\right)}\,.
\end{align}

We are interested in the effective dynamics of the electron, which for simplicity we assume to be along the $x$ axis, after taking into account its interaction with the  radiation field environment. To achieve that, we start by decomposing the phase-space weighing function $\exp{iS/\hbar}$ governing the full dynamics as $\exp{iS/\hbar}  = \exp{i(S_{\text{\tiny{S}}}[x]+S_{\text{\tiny{EM}}}[\mu]+S_{\mathrm{int}}[x,\Pi])/\hbar}$. 
Here, $S_{\text{\tiny{S}}}$ denotes the action corresponding to $\text{H}_{\text{\tiny{S}}}$,  $S_{\text{\tiny{int}}}[x,\Pi]$ is defined to be $S_{\text{\tiny{int}}}[x,\Pi]:=-e\int^{t}_{t_i} dt'x(t') \Pi_{\text{\tiny{E}}}\left(x(t'),t'\right)$ and $S_{\text{\tiny{EM}}}[\mu]:=S_{\text{\tiny{EM}}}[\mathrm{A}_{\perp},\Pi]$ inside the phase-space weighing function governs the time evolution of the free radiation field in which $\mu$ denotes the canonical degrees of freedom of the radiation field. In the light of the discussion around Eq.~\eqref{eq:TransitionMomentum}, with a slight abuse of notation, $\exp\lbrace\frac{i}{\hbar}S_{\text{\tiny{EM}}}\rbrace$ is understood to be simply the appropriate phase-space weighing function appearing inside the path integral with  $S_{\text{\tiny{EM}}}:=-\int^{t}_{t_i}d^3rdt'(\mathcal{H}_{\text{\tiny{EM}}}-\Pi \dot{\mathrm{A}}_{\perp})$ or $S_{\text{\tiny{EM}}}:=-\int^{t}_{t_i}d^3rdt'(\mathcal{H}_{\text{\tiny{EM}}}+\mathrm{A}_{\perp}\dot{\Pi})$ depending upon the basis states between which the transition amplitudes are calculated. In terms of these notations, the expression for the full system-environment density matrix $\hat{\rho}$ is given by
\begin{align}\label{eq:density_full}
&\bra{x^{\prime}_{\text{\tiny{f}}};\Pi^{f\prime}}\hat{\rho}(t)\ket{x_{\text{\tiny{f}}};\Pi^f} =\nonumber\\   
&\int_{\substack{{x(t) = x_{\text{\tiny{f}}} ,}\\ {x^{\prime}(t)= x^{\prime}_{\text{\tiny{f}}}}}}D[x,x^{\prime}]e^{\frac{i}{\hbar}(S_{\text{\tiny{S}}}[x^{\prime}]-S_{\text{\tiny{S}}}[x])}\rho^i_{\text{\tiny{S}}}\int_{\substack{{\Pi(t) = \Pi^f ,}\\ {\Pi^{\prime}(t)= \Pi^{ f\prime}}}}\left(D[\mu,\mu^{\prime}]\nonumber\right.\\
&\left.e^{\frac{i}{\hbar}(S_{\text{\tiny{EM}}}[\mu']+S_{\mathrm{int}}[x^{\prime},\Pi^{\prime}]-S_{\text{\tiny{EM}}}[\mu]-S_{\mathrm{int}}[x,\Pi])}\rho^{i}_{\text{\tiny{EM}}}\right)\,,
\end{align}
where $\ket{\Pi^f}$ denotes the (momentum) basis state of the environment \footnote{Note that the precise choice of the basis states is unimportant since the reduced density matrix is obtained after tracing over the environment} and
\begin{align}
\rho^i_{\text{\tiny{S}}}:=\rho_{\text{\tiny{S}}}(x^{\prime}_i,x_i,t_i)\,,\qquad \rho^{i}_{\text{\tiny{EM}}}:=\rho_{\text{\tiny{EM}}}(\Pi^{\prime}(t_i),\Pi(t_i),t_i)\,.
\end{align}
In writing Eq.~\eqref{eq:density_full}, we have also assumed the full density matrix
$\hat{\rho}(t_i)$ to be in the product state $\hat{\rho}(t_i) = \hat{\rho}_{\text{\tiny{S}}}(t_i)\otimes\hat{\rho}_{\text{\tiny{EM}}}(t_i)$ at the initial time $t_i$. 

To obtain the effective dynamics of the electron, we need to average over the radiation field environment. We notice that $S_{\text{\tiny{EM}}}[\mu]$ is quadratic in the environmental degrees of freedom while $S_{\mathrm{int}}[x,\Pi]$ is linear in both $x$ and $\Pi_{\text{\tiny{E}}}$. Therefore, after tracing over the environment, the integral involving the environmental degrees of freedom $\mu$ in Eq.~\eqref{eq:density_full}  yields a Gaussian in $x$ such that \cite{calzetta_hu_2008}
\begin{align}
&\int_{\mathrm{tr}}e^{\frac{i}{\hbar}(S^{\prime}_{\text{\tiny{EM}}}+S^{\prime}_{\mathrm{int}}-S_{\text{\tiny{EM}}}-S_{\mathrm{int}})}\rho_{\text{\tiny{EM}}}^{i} =\nonumber\\ &\quad\exp{\frac{i}{2\hbar}\iint dt_1dt_2 M_{ab}(t_1;t_2)x^a(t_1)x^b(t_2)}\,,\label{Mab}
\end{align}

where
\begin{align}
\int_{\mathrm{tr}}&:=\int d\Pi(t)\int_{\substack{{\Pi(t) = \Pi^{\prime}(t)}}}  D[\mu,\mu^{\prime}]\,,\nonumber\\ S^{\prime}_{\text{\tiny{EM}}}&:=S_{\text{\tiny{EM}}}[\mu']\,,\qquad
S^{\prime}_{\mathrm{int}}:=S_{\mathrm{int}}[x',\Pi']\,.
\end{align}
We have also introduced the vector notation with the convention $x^a=x$ for $a=1$, $x^a=x'$ for $a=2$ and $x_a=\eta_{ab}x^{b}$ with $\eta_{ab}=\mathrm{diag}(-1,1)$.

 It is the matrix $M_{ab}$ that determines the effective action of the system and contains the information about its interaction with the environment. These matrix elements can be obtained by acting with $\frac{\hbar}{i}\frac{\delta}{\delta x^{a}}\frac{\delta}{\delta x^{b}}|_{x^{a}=x^{b}=0}$ (where $x^a$ and $x^b$ are set to zero after taking the derivatives) on Eq.~\eqref{Mab} such that
\begin{align}
&M_{ab}(t_1;t_2)=\nonumber\\
&\quad\frac{ie^2}{\hbar}\int_{\mathrm{tr}}\Pi_{\text{\tiny{E}}a}\left(t_1\right) \Pi_{\text{\tiny{E}}b}\left(t_2\right) e^{\frac{i}{\hbar}(S_{\text{\tiny{EM}}}[\mu^{\prime}]-S_{\text{\tiny{EM}}}[\mu])}\rho_{\text{\tiny{EM}}}^{i}\,.
\end{align} 
Depending upon the value of the indices $a$ and $b$, the matrix elements correspond to the expectation values of the time-ordered ($\mathcal{T}$), anti-time ordered ($\tilde{\mathcal{T}}$), path-ordered or anti-path ordered products in the Heisenberg picture \cite{calzetta_hu_2008}. They are given by 
\begin{align}\label{eq:Mab}
&M_{ab}(t_1;t_2) =\nonumber\\ 
&\quad\frac{ie^2}{\hbar}\begin{bmatrix}
\left\langle\tilde{\mathcal{T}}\{ \hat{\Pi}_{\text{\tiny{E}}}(t_1) \hat{\Pi}_{\text{\tiny{E}}}(t_2)\}\right\rangle_0 & -\left\langle \hat{\Pi}_{\text{\tiny{E}}}(t_1) \hat{\Pi}_{\text{\tiny{E}}}(t_2)\right\rangle_0\\-\left\langle \hat{\Pi}_{\text{\tiny{E}}}(t_2) \hat{\Pi}_{\text{\tiny{E}}}(t_1)\right\rangle_0&\left\langle \mathcal{T}\{ \hat{\Pi}_{\text{\tiny{E}}}(t_1) \hat{\Pi}_{\text{\tiny{E}}}(t_2)\}\right\rangle_0 
\end{bmatrix}\,.
\end{align}
In Eq.~\eqref{eq:Mab}, the zero in the subscript denotes that the correlations of the environmental operator, i.e. the conjugate electric field operator $\hat{\Pi}_{\text{\tiny{E}}}$, are calculated by disregarding the system-environment interaction. Since the initial state of the environment is taken to be the vacuum state $\ket{0}$ of the radiation field, $\langle\cdot\rangle_0=\bra{0}\cdot\ket{0}$. Moreover, only the $x$ component of $\hat{\Pi}_{\text{\tiny{E}}}$ is understood to appear in Eq.~\eqref{eq:Mab} since the motion of the electron is taken to be along the $x$ axis for simplicity. Finally, since the electron is assumed to travel at nonrelativistic speeds $v\ll c$, we have also neglected the spatial dependence of $\hat{\Pi}_{\text{\tiny{E}}}$ inside the correlations (cf. Sec.~\ref{Sec:Kernels} for a more elaborate discussion of this approximation). 

Coming back to the density matrix, we see that after tracing over the environment in Eq.~\eqref{eq:density_full},  the reduced density matrix $\hat{\rho}_{r}$ of the electron that we are seeking takes the form
\begin{align}\label{eq:RedRho_formal}
&\bra{x^{\prime}_{\text{\tiny{f}}}}\hat{\rho}_{r}(t)\ket{x_{\text{\tiny{f}}}} =\nonumber\\ 
&\quad\int_{\substack{{x(t) = x_{\text{\tiny{f}}} ,}\\ {x^{\prime}(t)= x^{\prime}_{\text{\tiny{f}}}}}}D[x,x^{\prime}]e^{\frac{i}{\hbar}(S_{\text{\tiny{S}}}[x^{\prime}]-S_{\text{\tiny{S}}}[x]+S_{\text{\tiny{IF}}}[x,x^{\prime}])}\rho_r(x^{\prime}_i, x_i,t_i)\,,
\end{align}
where the so-called influence functional $S_{\text{\tiny{IF}}}$ \cite{Feynman_Vernon} is given by
\begin{align}\label{eq:InfluenceFuctionalx}
&S_{\text{\tiny{IF}}}[x,x^{\prime}] =\nonumber\\ 
&\quad\frac{ie^2}{2\hbar}\int_{t_i}^t dt_1dt_2\left[\left\langle\tilde{\mathcal{T}}\{ \hat{\Pi}_{\text{\tiny{E}}}(t_1) \hat{\Pi}_{\text{\tiny{E}}}(t_2)\}\right\rangle_0 x(t_1)x(t_2)\right.\nonumber\\ 
&\quad-\left\langle \hat{\Pi}_{\text{\tiny{E}}}(t_1) \hat{\Pi}_{\text{\tiny{E}}}(t_2)\right\rangle_0x(t_1)x^{\prime }(t_2)\nonumber\\
&\quad-\left\langle \hat{\Pi}_{\text{\tiny{E}}}(t_2) \hat{\Pi}_{\text{\tiny{E}}}(t_1)\right\rangle_0 x^{\prime }(t_1)x(t_2)\nonumber\\
&\quad\left.+\left\langle \mathcal{T}\{ \hat{\Pi}_{\text{\tiny{E}}}(t_1) \hat{\Pi}_{\text{\tiny{E}}}(t_2)\}\right\rangle_0x^{\prime }(t_1)x^{\prime }(t_2)\right]\,.
\end{align}
Here, the integral $\int_{t_i}^t$ stands for both the $t_1$ and the $t_2$ integrals, which run from $t_i$ to $t$. The influence functional $S_{\text{\tiny{IF}}}$ can also be written in the matrix notation as
\begin{align}
& S_{\text{\tiny{IF}}}[x,x^{\prime}] =\nonumber\\ 
&\quad\frac{1}{2}\int_{t_i}^{t} dt_1dt_2\begin{bmatrix}
x(t_1) &x^{\prime}(t_1)
\end{bmatrix}\cdot\begin{bmatrix}
M_{11}&M_{12}\\
M_{21}&M_{22}
\end{bmatrix}\cdot\begin{bmatrix}
x(t_2)\\x^{\prime}(t_2)
\end{bmatrix}\,.
\end{align}
As it is more convenient, we make a change of basis to $(X\,,u)$  defined by
\begin{align}
X(t):=&(x^{\prime}(t)+x(t))/2\,,\quad u(t)=x^{\prime}(t)-x(t)\,,
\end{align}
in which the influence functional transforms as
\begin{align}
&S_{\text{\tiny{IF}}}[X,u] =\nonumber\\ 
&\quad\frac{1}{2}\int_{t_i}^{t} dt_1dt_2\begin{bmatrix}
X(t_1) &u(t_1)
\end{bmatrix}
\cdot\begin{bmatrix}
\tilde{M}_{11}&\tilde{M}_{12}\\
\tilde{M}_{21}&\tilde{M}_{22}
\end{bmatrix}\cdot\begin{bmatrix}
X(t_2)\\u(t_2)
\end{bmatrix}\,,
\end{align}
with
\begin{align}
\tilde{M}_{11}&=M_{11}+M_{12}+M_{21}+M_{22}\,,\nonumber\\
\tilde{M}_{12}&=\frac{1}{2}\left((M_{12}-M_{21})+(M_{22}-M_{11})\right)\,,\nonumber\\
\tilde{M}_{21}&=\frac{1}{2}\left(-(M_{12}-M_{21})+(M_{22}-M_{11})\right)\,,\nonumber\\
\tilde{M}_{22}&=\frac{1}{4}((M_{11}+M_{22})-(M_{12}+M_{21}))\,.
\end{align}
Further, with the help of Eq.~\eqref{eq:Mab}, we get the following relations:
\begin{align}
M_{11}+M_{22}&=-(M_{12}+M_{21})\nonumber\\
&=\frac{ie^2}{\hbar}\left\langle\{ \hat{\Pi}_{\text{\tiny{E}}}(t_1), \hat{\Pi}_{\text{\tiny{E}}}(t_2)\}\right\rangle_0\,,\\
M_{12}-M_{21}&=\frac{ie^2}{\hbar}\left\langle\left[ \hat{\Pi}_{\text{\tiny{E}}}(t_2), \hat{\Pi}_{\text{\tiny{E}}}(t_1)\right]\right\rangle_0\,,\\ M_{22}-M_{11}&=\frac{ie^2}{\hbar}\left\langle\left[ \hat{\Pi}_{\text{\tiny{E}}}(t_1), \hat{\Pi}_{\text{\tiny{E}}}(t_2)\right]\right\rangle_0\mathrm{sgn}(t_1-t_2)\,.
\end{align}
Using these relations, the matrix $\tilde{M}$ can be written as  
\begin{align}\label{eq:thetafunction}
\tilde{M}_{11}&=0\,,\nonumber\\
\tilde{M}_{12}&=\frac{ie^2}{\hbar}\left\langle\left[ \hat{\Pi}_{\text{\tiny{E}}}(t_2), \hat{\Pi}_{\text{\tiny{E}}}(t_1)\right]\right\rangle_0\theta(t_2-t_1)\,,\nonumber\\
\tilde{M}_{21}&=\frac{ie^2}{\hbar}\left\langle\left[ \hat{\Pi}_{\text{\tiny{E}}}(t_1), \hat{\Pi}_{\text{\tiny{E}}}(t_2)\right]\right\rangle_0\theta(t_1-t_2)\,,\nonumber\\
\tilde{M}_{22}&=\frac{ie^2}{2\hbar}\left\langle\{ \hat{\Pi}_{\text{\tiny{E}}}(t_1), \hat{\Pi}_{\text{\tiny{E}}}(t_2)\}\right\rangle_0\,,
\end{align}
where $\theta(t)$ is the Heaviside step function. 
Thus, in the $(X,u)$ basis, the influence functional in Eq.~\eqref{eq:InfluenceFuctionalx} takes the compact form 
\begin{align}\label{eq:InfluenceFunctional}
&S_{\text{\tiny{IF}}}[X,u](t) =\nonumber\\ 
&\int_{t_i}^t dt_1dt_2 \left[i\frac{ u(t_1)\mathcal{N}(t_1;t_2) u(t_2)}{2} + u(t_1)\mathcal{D}(t_1;t_2)X(t_2)\right]\,,
\end{align}
where the noise kernel $\mathcal{N}$ and the dissipation kernel  $\mathcal{D}$ are defined as
\begin{align}
\mathcal{N}(t_1;t_2):=&\frac{e^2}{2\hbar}\left\langle\{ \hat{\Pi}_{\text{\tiny{E}}}(t_1), \hat{\Pi}_{\text{\tiny{E}}}(t_2)\}\right\rangle_0\,,\nonumber\\
\mathcal{D}(t_1;t_2):=&\frac{ie^2}{\hbar}\left\langle\left[ \hat{\Pi}_{\text{\tiny{E}}}(t_1), \hat{\Pi}_{\text{\tiny{E}}}(t_2)\right]\right\rangle_0\theta(t_1-t_2)\,.
\end{align} 
Having determined the full effective action for the electron in terms of the influence functional, the master equation for its reduced density matrix in Eq.~\eqref{eq:RedRho_formal} can now be derived. From Eq.~\eqref{eq:RedRho_formal}, it can be seen that the time derivative of the reduced density matrix will have, in addition to the standard Liouville--von Neumann term, the contribution coming from the influence functional. In order to compute that, the rate of change of $S_{\text{\tiny{IF}}}$ needs to be evaluated. It is given by
\begin{align}\label{eq:IF_Time_Derivative}
&\delta_t S_{\text{\tiny{IF}}}[X,u]=\nonumber\\ 
&\quad u(t)\int_{t_i}^t dt_1 \left(i \mathcal{N}(t;t_1)u(t_1)+\mathcal{D}(t;t_1)X(t_1)\right)\,.
\end{align}
Using Eq.~\eqref{eq:IF_Time_Derivative}, in terms of the original $(x,x^{\prime})$ basis, the master equation can now be written as
\begin{widetext}
\begin{align}\label{eq:SecondOrderApprox}
\partial_t\rho_r(x^{\prime}_{\text{\tiny{f}}},x_{\text{\tiny{f}}},t)&=-\frac{i}{\hbar}\bra{x'_{\text{\tiny{f}}}}\left[\hat{\mathrm{H}}_s,\hat{\rho}_{r}\right]\ket{x_{\text{\tiny{f}}}} + \frac{i}{\hbar}\int_{\substack{{x(t) = x_{\text{\tiny{f}}} ,}\\ {x^{\prime}(t)= x^{\prime}_{\text{\tiny{f}}}}}}D[x,x^{\prime}]\delta_t S_{\text{\tiny{IF}}}[x^{\prime},x]e^{\frac{i}{\hbar}(S_{\text{\tiny{S}}}[x^{\prime}]-S_{\text{\tiny{S}}}[x]+S_{\text{\tiny{IF}}}[x,x^{\prime}])}\rho_r(x'_i,x_i,t_i)\nonumber\\
&\approx-\frac{i}{\hbar}\bra{x^{\prime}_{\text{\tiny{f}}}}\left[\hat{\mathrm{H}}_s,\hat{\rho}_{r}\right]\ket{x_{\text{\tiny{f}}}}+\frac{i}{\hbar}\int_{\substack{{x(t) = x_{\text{\tiny{f}}} ,}\\ {x^{\prime}(t)= x^{\prime}_{\text{\tiny{f}}}}}}D[x,x^{\prime}]\delta_t S_{\text{\tiny{IF}}}[x^{\prime},x]e^{\frac{i}{\hbar}(S_{\text{\tiny{S}}}[x^{\prime}]-S_{\text{\tiny{S}}}[x])}\rho_r(x'_i,x_i,t_i)\nonumber\\
&\approx-\frac{i}{\hbar}\bra{x^{\prime}_{\text{\tiny{f}}}}\left[\hat{\mathrm{H}}_s,\hat{\rho}_{r}\right]\ket{x_{\text{\tiny{f}}}}\nonumber\\
&\hspace{11pt}-\frac{1}{\hbar}(x^{\prime}_{\text{\tiny{f}}}-x_{\text{\tiny{f}}})\int_{t_i}^t dt_1\mathcal{N}(t;t_1)\int_{\substack{{x(t) = x_{\text{\tiny{f}}} ,}\\ {x^{\prime}(t)= x^{\prime}_{\text{\tiny{f}}}}}}D[x,x^{\prime}](x^{\prime}(t_1)-x(t_1))  e^{\frac{i}{\hbar}(S_{\text{\tiny{S}}}[x^{\prime}]-S_{\text{\tiny{S}}}[x])}\rho_r(x'_i,x_i,t_i)\nonumber\\
&\hspace{11pt}+\frac{i}{2\hbar}(x^{\prime}_{\text{\tiny{f}}}-x_{\text{\tiny{f}}})\int_{t_i}^t dt_1\mathcal{D}(t;t_1)\int_{\substack{{x(t) = x_{\text{\tiny{f}}} ,}\\ {x^{\prime}(t) =x^{\prime}_{\text{\tiny{f}}}}}}D[x,x^{\prime}](x^{\prime}(t_1)+x(t_1))  e^{\frac{i}{\hbar}(S_{\text{\tiny{S}}}[x^{\prime}]-S_{\text{\tiny{S}}}[x])}\rho_r(x'_i,x_i,t_i)\,.
\end{align}
\end{widetext}
For the second term on the right-hand side in the second line of Eq.~\eqref{eq:SecondOrderApprox}, $S_{\text{\tiny{IF}}}$ has been omitted in the exponential. This is because $S_{\text{\tiny{IF}}}$ is second order in the coupling constant and is already present adjacent to the exponential. Since the calculations are limited to second order in the interactions, $S_{\text{\tiny{IF}}}$ can be neglected inside the exponential. 

To simplify the master equation further, we note that the last two lines of Eq.~\eqref{eq:SecondOrderApprox} can be written much more compactly. This is because \cite{calzetta_hu_2008}
\begin{align}\label{eq:IdentityOne}
&\int_{\substack{{x(t) = x_{\text{\tiny{f}}} ,}\\{x^{\prime}(t)= x^{\prime}_{\text{\tiny{f}}}}}}D[x,x^{\prime}]x^{\prime}(t_1)  e^{\frac{i}{\hbar}(S_{\text{\tiny{S}}}[x^{\prime}]-S_{\text{\tiny{S}}}[x])}\rho_r(x'_i,x_i,t_i)=\nonumber\\ 
&\quad \int dx^{\prime}(t_1)\bra{x^{\prime}_{\text{\tiny{f}}}}\hat{U}_s(t;t_1)\ket{x^{\prime}(t_1)}x^{\prime}(t_1)\times\nonumber\\
&\quad\times\bra{x^{\prime}(t_1)}\hat{U}_s(t_1;t_i)\hat{\rho}_r(t_i)\hat{U}_s^{-1}(t;t_i)\ket{x_{\text{\tiny{f}}}}\nonumber\\
&=\bra{x^{\prime}_{\text{\tiny{f}}}}\hat{U}_s(t;t_1)\hat{x}\hat{U}_s(t_1;t_i)\hat{\rho}_r(t_i)\hat{U}_s^{-1}(t;t_i)\ket{x_{\text{\tiny{f}}}}\nonumber\\
&=\bra{x^{\prime}_{\text{\tiny{f}}}}\hat{U}_s(t;t_1)\hat{x}\hat{U}_s(t_1;t_i)\hat{U}_s^{-1}(t;t_i)\times\nonumber\\
&\quad\times\hat{U}_s(t;t_i)\hat{\rho}_r(t_i)\hat{U}_s^{-1}(t;t_i)\ket{x_{\text{\tiny{f}}}}\nonumber\\
&=\bra{x^{\prime}_{\text{\tiny{f}}}}\hat{U}_s(t;t_1)\hat{x}\hat{U}_s^{-1}(t;t_1)\hat{\rho}_r(t)\ket{x_{\text{\tiny{f}}}} \nonumber\\
&=\bra{x^{\prime}_{\text{\tiny{f}}}}\hat{x}_{\text{\tiny{H}}_s}(-\tau)\hat{\rho}_r(t)\ket{x_{\text{\tiny{f}}}}\,,
\end{align}
where
\begin{align}\label{eq:xH}
\hat{x}_{\text{\tiny{H}}_s}(-\tau):= \hat{U}^{-1}_s(t-\tau;t)\hat{x}\hat{U}_s(t-\tau;t)\,,\qquad \tau:=t-t_1\,.
\end{align}
Here, $\hat{U}_s(t-\tau;t)$ is the unitary operator that evolves the state vector of the system from time $t$ to $t-\tau$ via the system Hamiltonian $\hat{\mathrm{H}}_s$, and the operator $\hat{x}$ without the subscript is the usual Schr\"{o}dinger operator such that 
\begin{align}
\hat{x}_{\text{\tiny{H}}_s}(0) = \hat{x}\,.
\end{align} 
Similarly, we also have the analogous relation 
\begin{align}\label{eq:IdentityTwo}
&\int_{\substack{{x(t) = x_{\text{\tiny{f}}} ,}\\{x^{\prime}(t)= x^{\prime}_{\text{\tiny{f}}}}}}D[x,x^{\prime}]x(t_1)  e^{\frac{i}{\hbar}(S_{\text{\tiny{S}}}[x^{\prime}]-S_{\text{\tiny{S}}}[x])}\rho_r(x'_i,x_i,t_i) =\nonumber\\ &\quad\bra{x^{\prime}_{\text{\tiny{f}}}}\hat{\rho}_r(t)\hat{x}_{\text{\tiny{H}}_s}(-\tau)\ket{x_{\text{\tiny{f}}}}\,.
\end{align}	
Using Eqs.~\eqref{eq:IdentityOne}--\eqref{eq:IdentityTwo} and replacing the $t_1$ integral  with the $\tau$ integral $(t_1 = t-\tau)$, the master equation~\eqref{eq:SecondOrderApprox} takes the compact form
\begin{align}\label{eq:TimeEvint}
&\partial_t\rho_r(x^{\prime}_{\text{\tiny{f}}},x_{\text{\tiny{f}}},t)=\nonumber\\
&\quad-\frac{i}{\hbar}\bra{x^{\prime}_{\text{\tiny{f}}}}\left[\hat{\mathrm{H}}_s,\hat{\rho}_{r}(t)\right]\ket{x_{\text{\tiny{f}}}}+\frac{(x^{\prime}_{\text{\tiny{f}}}-x_{\text{\tiny{f}}})}{\hbar}\int_{0}^{t-t_{i}} d\tau\left(\right.\nonumber\\
&\quad-\mathcal{N}(t;t-\tau)\bra{x^{\prime}_{\text{\tiny{f}}}}\left[\hat{x}_{\text{\tiny{H}}_s}(-\tau),\hat{\rho}_r(t)\right]\ket{x_{\text{\tiny{f}}}}\nonumber\\
&\quad\left.+\frac{i}{2}\mathcal{D}(t;t-\tau)\bra{x^{\prime}_{\text{\tiny{f}}}}\{\hat{x}_{\text{\tiny{H}}_s}(-\tau),\hat{\rho}_r(t)\}\ket{x_{\text{\tiny{f}}}}\right)\,.
\end{align}
The eigenvalues outside of the integrals in Eq.~\eqref{eq:TimeEvint} can be obtained by acting with the position operator $\hat{x}$ such that
\begin{align}\label{eq:TimeEvint2}
&\bra{x^{\prime}_{\text{\tiny{f}}}}\partial_t\hat{\rho}_r\ket{x_{\text{\tiny{f}}}}=\nonumber\\
&-\frac{i}{\hbar}\bra{x^{\prime}_{\text{\tiny{f}}}}\left[\hat{\mathrm{H}}_s,\hat{\rho}_{r}(t)\right]\ket{x_{\text{\tiny{f}}}} \nonumber\\
&-\frac{1}{\hbar}\int_{0}^{t-t_i} d\tau \mathcal{N}(t;t-\tau)\bra{x^{\prime}_{\text{\tiny{f}}}}\left[\hat{x},\left[\hat{x}_{\text{\tiny{H}}_s}(-\tau),\hat{\rho}_{r}(t)\right]\right]\ket{x_{\text{\tiny{f}}}}\nonumber\\
&+\frac{i}{2\hbar}\int_{0}^{t-t_i}
d\tau\mathcal{D}(t;t-\tau)\bra{x^{\prime}_{\text{\tiny{f}}}}\left[\hat{x},\{\hat{x}_{\text{\tiny{H}}_s}(-\tau),\hat{\rho}_{r}(t)\}\right]\ket{x_{\text{\tiny{f}}}}\,.
\end{align}
We can now write the master equation for the reduced density matrix of the nonrelativistic electron, up to second order in the interactions, in a basis-independent operator form. It reads
\begin{align}\label{eq:MasterEquation}
\partial_t\hat{\rho}_r=&-\frac{i}{\hbar}\left[\hat{\mathrm{H}}_s,\hat{\rho}_{r}(t)\right]\nonumber\\  &-\frac{1}{\hbar}\int_{0}^{t-t_i} d\tau \mathcal{N}(t;t-\tau)\left[\hat{x},\left[\hat{x}_{\text{\tiny{H}}_s}(-\tau),\hat{\rho}_{r}(t)\right]\right]\nonumber\\&+\frac{i}{2\hbar}\int_{0}^{t-t_i}
d\tau\mathcal{D}(t;t-\tau)\left[\hat{x},\{\hat{x}_{\text{\tiny{H}}_s}(-\tau),\hat{\rho}_{r}(t)\}\right]\,.
\end{align}
The first line of the master equation is the usual Liouville--von Neumann evolution and involves only the system Hamiltonian $\hat{\mathrm{H}}_s$, while the second and the third lines explicitly encode the system's interaction with the environment. We remember that due to the coupling between the position of the electron and the transverse electric field in Eq.~\eqref{eq:Lag}, the system Hamiltonian receives an additional contribution such that $\hat{\mathrm{H}}_s=\hat{p}^2/(2m)+\hat{V}_0+\hat{V}_{\text{\tiny{EM}}}$, where, having introduced a UV cutoff in the calculations and considering the motion of the electron along the $x$ axis only, $\hat{V}_{\text{\tiny{EM}}}(x)= \frac{e^2\omega^3_{\text{\tiny{max}}}}{3\pi^2\epsilon_0c^3}\hat{x}^2$ (cf. Sec.~\ref{Sec:Hamiltonian}).
Moreover, since the master equation is valid up to second order in the interactions and since the operator $\hat{x}_{\text{\tiny{H}}_s}(-\tau)$ appears alongside the dissipation and the noise kernels (which are already second order in $e$), the time evolution governed by $\hat{U}_s(t-\tau;t)$ in Eq.~\eqref{eq:xH} is understood to involve only $\hat{V}_0$ and not $\hat{V}_{\text{\tiny{EM}}}$. Therefore, up to second order in the interactions, $\hat{V}_{\text{\tiny{EM}}}$ contributes to the master equation only via the Liouville--von Neumann term.

\section{The dissipation and the noise kernels}\label{Sec:Kernels}
Having obtained the formal expression of the master equation \eqref{eq:MasterEquation}, we now proceed towards calculating the kernels (noise and dissipation) explicitly.  In order to do so, we remember that only the $x$ component of $\hat{\Pi}_{\text{\tiny{E}}}$ is relevant for the kernels.  Its expression,
upon standard quantization, in terms of the creation and annihilation operators and the $x$ component of the unit polarization vector ${\varepsilon}^x_{\k}$, is given by \cite{TannoudjiPartOneChapterThree}
\begin{align}\label{eq:quantized_efield}
\hat{\Pi}_{\text{\tiny{E}}}(\r,t) ={}&iC\int d^3k \sqrt{k}\sum_{\textbf{$\varepsilon$}}\hat{a}_{\varepsilon}(\k)e^{i(\k\cdot \r-\omega t)}\varepsilon^x_{\k}+\mathrm{c.c}\,,\nonumber\\
C:={}&\left(\frac{\hbar c}{2\epsilon_{0} (2\pi)^3}\right)^{\frac{1}{2}}\,.
\end{align}

Using Eq.~\eqref{eq:quantized_efield}, we obtain the expression for the vacuum expectation value of the two-point correlator to be 
\begin{align}\label{eq:correlator_integral}
&\bra{0}{ \hat{\Pi}_{\text{\tiny{E}}}}(x(t_1),t_1){ \hat{\Pi}_{\text{\tiny{E}}}}(x(t_2),t_2)\ket{0} = \nonumber\\
&\quad\frac{-i\hbar c}{2\epsilon_{0}4\pi^2}\hat\Box\left\{\frac{1}{r}\int_{0}^{\infty} dk e^{-ikc\tau}\left(e^{ikr} - e^{-i k r}\right)\right\}\,,
\end{align}
with $\tau:=t_1-t_2$ and 
\begin{align}
r:=|x(t_1)-x(t_2)|\,,\qquad \hat\Box:= -\frac{1}{c^2}\partial^2_{\tau}+\partial^2_{r}\,.
\end{align}
The evaluation of the integral in Eq.~\eqref{eq:correlator_integral} requires the introduction of a UV cutoff, as it was needed, for example, in Sec.~\ref{Sec:Hamiltonian}. For that, as in Sec.~\ref{Sec:Hamiltonian}, we resort to the standard Hadamard finite part prescription \cite{calzetta_hu_2008} in which the convergence factor $e^{-\omega_{k}/\omega_{\text{\tiny{max}}}}$  (with $\omega_{k} = kc$) is introduced inside the integral.  Physically, this prescription cuts off the contribution coming from the modes $\omega_{k}\gg\omega_{\text{\tiny{max}}}$ and, mathematically, it is the same as using the $i \epsilon$ prescription where one sends $\tau\to \tau-i\epsilon$, with 
\begin{align}
\epsilon = \frac{1}{\omega_{\text{\tiny{max}}}}\,.
\end{align}
Evaluating the integral by using this prescription we get
\begin{align}\label{eq:field_correlator_spatial}
\bra{0}{ \hat{\Pi}_{\text{\tiny{E}}}}(1){ \hat{\Pi}_{\text{\tiny{E}}}}(2)\ket{0}  &={}  \frac{\hbar c}{4\pi^2\epsilon_{0}}\hat\Box\left\{\frac{1}{r^2 - c^2 (\tau-i\epsilon)^2}\right\}\nonumber\\ 
&={}\frac{\hbar c}{\pi^2\epsilon_{0}}\frac{1}{\left(r^2 - c^2 (\tau-i\epsilon)^2\right)^2}\,.
\end{align}
For the correlator in  Eq.~\eqref{eq:field_correlator_spatial}, we can ignore the spatial dependence since $r\ll c\tau$ for the electron traveling at speeds $v\ll c$. In this limit, the correlator becomes
\begin{align}\label{eq:field_correlator}
\bra{0} \hat{\Pi}_{\text{\tiny{E}}}(1) \hat{\Pi}_{\text{\tiny{E}}}(2)\ket{0} \approx \frac{\hbar}{\pi^2\epsilon_{0}c^3\left(\tau-i\epsilon\right)^4}\,.
\end{align}
Using Eq.~\eqref{eq:field_correlator}, we obtain the explicit functional form of the  kernels to be
\begin{align}
\mathcal{N}(\tau) &= \frac{e^2}{\pi^2\epsilon_{0}c^3}\frac{\left(\epsilon^4-6\epsilon^2\tau^2+\tau^4\right)}{\left(\epsilon^2+\tau^2\right)^4}\,,\label{eq:NoiseKernel}\\ 
\mathcal{D}(\tau) &=\frac{8 e^2}{\pi^2\epsilon_{0}c^3}\frac{\epsilon\tau(\epsilon^2-\tau^2)}{(\epsilon^2+\tau^2)^4}\theta(\tau)\label{eq:DissipationKernelOne}\,.
\end{align}
Further, with some algebraic manipulation, the dissipation kernel can be expressed more compactly as
\begin{align}
\mathcal{D}(\tau) =  \frac{ e^2}{3\pi^2\epsilon_{0}c^3}\theta(\tau)\frac{d^3}{d\tau^3}\left(\frac{\epsilon}{\tau^2+\epsilon^2}\right)\,.
\end{align}
Noticing that
\begin{align}\label{eq:deltafunction}
\frac{\epsilon}{\tau^2+\epsilon^2} = \frac{d}{d\tau}\tan^{-1}(\tau/\epsilon) = \pi \delta_{\epsilon}(\tau)\,,
\end{align}
we arrive at the expression
\begin{align}\label{eq:DissipationKernel}
\mathcal{D}(\tau) = \frac{e^2}{3\pi\epsilon_{0}c^3}\theta(\tau)\frac{d^3}{d\tau^3}\delta_{\epsilon}(\tau)\,.
\end{align}
The last equality in Eq.~\eqref{eq:deltafunction} can be understood in the limit $\epsilon\to 0$, in which the function $\tan^{-1}(\tau/\epsilon)$ takes the shape of a step function such that its time derivative approaches the Dirac delta $\delta_{\epsilon}(\tau)$.

It is important to emphasize the context in which the so-called dipole approximation is applied in our work. Strictly speaking, the approximation we make is only to ignore the spatial variation of the two-point correlations in going from Eq.~\eqref{eq:field_correlator_spatial} to  Eq.~\eqref{eq:field_correlator}. This approximation remains valid as long as the electron travels at speeds $v\ll c$. This is conceptually different from the more common application of the standard dipole approximation. There, one similarly ignores the spatial variation of the radiation field,  but in the light of the particle being trapped  over length scales that are much shorter than the characteristic wavelengths of the  radiation field. However, this is not the physical situation that our calculations are restricted to. It is clear that in only ignoring the spatial variation of the two-point correlations, we also describe the physical situation of a freely moving charged particle, not necessarily confined around the origin, as long as it moves at speeds $v\ll c$.

\section{The equation of motion}\label{Sec:EOM} 
Using the master equation~\eqref{eq:MasterEquation}, we can obtain the coupled equations for the time evolution of $\langle\hat{x}\rangle$ and $\langle{\hat{p}}\rangle$. It is interesting to compare the quantum mechanical EOM with the one derived classically.

Within classical electrodynamics, a charged spherical shell of radius $\mathrm{R}$ which is accelerated by an external force $\mathrm{F}_{\mathrm{ext}}$, experiences an extra recoil force (radiation reaction) due to the emission of radiation. By taking the limit $\mathrm{R}\to 0$ in the equation describing its dynamics, one obtains the Abraham-Lorentz equation,
\begin{align}\label{eq:AL}
m_{\text{\tiny{R}}}\ddot{x} = \mathrm{F}_{\mathrm{ext}}+\frac{2\hbar\alpha}{3c^2}\dddot{x}\,,
\end{align} 
where $m_{\text{\tiny{R}}}$ denotes the observed renormalized mass. See, for example, Refs.~\cite{Pearle1982,Griffiths2010} and the references therein for the derivation of the AL equation.  The triple derivative term appearing in Eq.~\eqref{eq:AL} can be interpreted as the friction term that leads to energy loss due to radiation emission. For instance, when the external potential is taken to be $V_0(x) = (1/2)m\omega^2_0x^2$, one has   $\dddot{x}\approx-\omega_{0}^2\dot{x}$ \cite{TannoudjiPartOneChapterOne}. However, the issue with Eq.~\eqref{eq:AL} is that the same triple derivative term persists even when the external potential is switched off, leading to an exponential increase of the particle's acceleration. A discussion of the AL equation and the problems associated with it can be found in \cite{Coleman1982, Pearle1982, Griffiths2017, Griffiths2010} and the references therein; Ref.~\cite{spohn_2004} offers an elaborate and mathematically accurate review and discussion on the subject.

In the case that we are considering, the rate of change of the expectation values is calculated from Eq.~$\eqref{eq:MasterEquation}$. The coupled differential equations for $\langle\hat{x}\rangle$ and $\langle{\hat{p}}\rangle$ are given by (cf. Appendix~\ref{App:AppendixB})
\begin{align}
\frac{d}{dt}\langle\hat{x}\rangle =& \mathrm{Tr}(\hat{x}\dot{\hat{\rho}}_{r})= \frac{\langle\hat{p}\rangle}{m}\,,\label{eq:AvX}\\
\frac{d}{dt}\langle\hat{p}\rangle =& -\langle\hat{V}_0,_{{x}}\rangle+\mathrm{Tr}\left(\hat{\rho}_{r}(t)\int_{0}^{t-t_i}d\tau \mathcal{D}(\tau)\hat{x}_{\text{\tiny{H}}_s}(-\tau)\right)\nonumber\\
&-\frac{2e^2\omega^3_{\text{\tiny{max}}}\langle\hat{x}\rangle}{3\pi^2\epsilon_0c^3}\label{eq:AvP1}\,.
\end{align}
While it might not be apparent at the first glance, Eq.~\eqref{eq:AvP1} is actually local in time due the form of $\mathcal{D}(\tau)$ in Eq.~\eqref{eq:DissipationKernel}. To see this explicitly, the integral involving the dissipation kernel needs to be evaluated. We do so by integrating by parts such that the derivatives acting on $\delta_{\epsilon}$ [appearing in the expression for $\mathcal{D}(\tau)$ in Eq.~\eqref{eq:DissipationKernel}] are shifted onto the adjacent function. We calculate the integral explicitly in Appendix~\ref{App:AppendixA} and derive the following identity:
\begin{align}\label{eq:integral_dissipation_kernel}
\int_{0}^td\tau\mathcal{D}(\tau)f(\tau) = &-\frac{2\alpha\hbar}{3c^2} f'''(0) -\frac{4\alpha\hbar\omega_{\text{\tiny{max}}}}{3\pi c^2} f''(0)\nonumber\\
&+\frac{2e^2\omega^3_{\text{\tiny{max}}}f(0)}{3\pi^2\epsilon_0c^3}\,.
\end{align}
Here, the prime denotes the derivative taken with respect to $\tau$ and $\alpha=e^2/(4\pi\epsilon_0\hbar c)$ is the fine-structure constant.
Using identity~\eqref{eq:integral_dissipation_kernel}, Eq.~\eqref{eq:AvP1} becomes
\begin{align}
\frac{d}{dt}\langle\hat{p}\rangle = &-\langle\hat{V_0},_{{x}}\rangle-\frac{4\alpha\hbar\omega_{\text{\tiny{max}}}}{3\pi c^2}\mathrm{Tr}\left(\hat{\rho}_{r}(t)\left.\frac{d^2}{d\tau^2}\hat{x}_{\text{\tiny{H}}_s}(-\tau)\right|_{\tau=0}\right)\nonumber\\
&-\frac{2\alpha\hbar}{3c^2}\mathrm{Tr}\left(\hat{\rho}_{r}(t)\left.\frac{d^3}{d\tau^3}\hat{x}_{\text{\tiny{H}}_s}(-\tau)\right|_{\tau=0}\right)\,.\label{eq:AvP2}
\end{align}
We see that in the EOM~\eqref{eq:AvP2}, only the original bare potential $\hat{V}_0$ remains because the contribution coming from $\hat{V}_{\text{\tiny{EM}}}$ in the second line of Eq.~\eqref{eq:AvP1} is canceled by the term in the second line of the integral~\eqref{eq:integral_dissipation_kernel}, after one introduces the cutoff consistently throughout the calculations. For more details, we refer to Appendices~\ref{App:AppendixA} and~\ref{App:AppendixB}. The same cancellation was also argued for in the quantum Langevin equation derived in \cite{Caldeira1991}. However, there it was argued that this cancellation occurs after one assumes a specific model for the charge distribution of the electron. In this work, we show that up to second order in the interactions, the cancellation occurs for any value of the cutoff $\omega_{\text{\tiny{max}}}$, as long as it is introduced consistently throughout the calculations and without making any additional assumptions concerning the charge distribution of the electron.

The time derivatives of $\hat{x}_{\text{\tiny{H}}_s}$ in Eq.~\eqref{eq:AvP2} can be easily computed, since from Eq.~\eqref{eq:xH} we have the relation [discarding $\hat{V}_{\text{\tiny{EM}}}$ up to second order in Eq.~\eqref{eq:AvP2}]
\begin{align}\label{eq:DerivativeXH}
\frac{d}{d\tau}\hat{x}_{\text{\tiny{H}}_s}(-\tau) = -\frac{i}{\hbar}\left[\hat{V}_0 +\frac{\hat{p}^2}{2m},\hat{x}_{\text{\tiny{H}}_s}(-\tau)\right]\,.
\end{align}
First we consider the situation when the external potential is switched off. From Eq.~\eqref{eq:DerivativeXH}, with $\hat{V}_0 =0$, taking another time derivative of $\hat{x}_{\text{\tiny{H}}_s}$ we get 
\begin{align}
\left.\frac{d^2}{d\tau^2}\hat{x}_{\text{\tiny{H}}_s}(-\tau)\right|_{\tau=0}&=\left(\frac{-i}{\hbar}\right)^2\left[\frac{\hat{p}^2}{2m},\left[\frac{\hat{p}^2}{2m},\hat{x}\right]\right]=0\,,\label{eq:DoubleCommutator}
\end{align}
where, in Eq.~\eqref{eq:DoubleCommutator}, we have also used the relation $\hat{x}_{\text{\tiny{H}}_s}(0)=\hat{x}$. Similarly, the third derivative term appearing in Eq.~\eqref{eq:AvP2} also vanishes. Therefore, when $\hat{V}_0=0$, Eq.~\eqref{eq:AvP2} simply reduces to 
\begin{align}\label{eq:FreeParticle}
\frac{d}{dt}\langle\hat{p}\rangle = 0\,. 
\end{align}
Unlike the AL equation~\eqref{eq:AL}, we see that up to second order in the interactions, there are no solutions which allow for an exponential increase of the particle's acceleration in the absence of an external potential.

Next we consider the case $\hat{V}_0\neq 0$. When the potential does not explicitly depend on time, such that $\hat{V}_0 =\hat{V}_0(x)$, the double and triple derivative terms in Eq.~\eqref{eq:AvP2} yield double and triple commutators with respect to the system Hamiltonian (discarding $\hat{V}_{\text{\tiny{EM}}}$) respectively. Equation~\eqref{eq:AvP2} can then be written as
\begin{align}\label{eq:AvPTrace}
\frac{d}{dt}\langle\hat{p}\rangle ={}&\mathrm{F}_{\mathrm{ext}}
+\frac{4\alpha\hbar\omega_{\text{\tiny{max}}}}{3\pi c^2}\mathrm{Tr}\left(\frac{1}{\hbar^2}\hat{\rho}_{r}(t) \left[\hat{\mathrm{H}}_s,\left[\hat{\mathrm{H}}_s,\hat{x}\right]\right]\right)\nonumber\\
&-\frac{2\alpha\hbar}{3c^2}\mathrm{Tr}\left(\frac{i}{\hbar^3}\hat{\rho}_{r}(t) \left[\hat{\mathrm{H}}_s,\left[\hat{\mathrm{H}}_s,\left[\hat{\mathrm{H}}_s,\hat{x}\right]\right]\right]\right)\,.
\end{align}
Here, we have defined $\mathrm{F}_{\mathrm{ext}}:=-\langle\hat{V_0}(x),_x\rangle$. Due to the presence of $\hat{V}_0(x)$, the commutators of $\hat{\mathrm{H}}_s$ with $\hat{x}$ no longer vanish. To simplify the equation further, we shift the commutators onto the density matrix using the cyclic property $\mathrm{Tr}(\hat{a}\cdot[\hat{b},\hat{c}]) = \mathrm{Tr}([\hat{a},\hat{b}]\cdot\hat{c})
$ such that
\begin{align}\label{eq:cyclic}
\mathrm{Tr}\left(\hat{\rho}_{r} \left[\hat{\mathrm{H}}_s,\left[\hat{\mathrm{H}}_s,\hat{x}\right]\right]\right) &= \mathrm{Tr}\left(\hat{x} \left[\hat{\mathrm{H}}_s,\left[\hat{\mathrm{H}}_s,\hat{\rho}_r\right]\right]\right)\,.
\end{align}
The same relationship also holds for the triple commutator term, with an additional minus sign.  Remembering that the master equation is only valid up to second order in the interaction, it is sufficient to evaluate the trace in Eq.~\eqref{eq:AvPTrace} at zeroth order. This implies that within the trace, the time dependence of the density matrix can be evaluated only by retaining the Liouville--von Neumann term in Eq.~\eqref{eq:MasterEquation}. The right-hand side of Eq.~\eqref{eq:cyclic} thus becomes proportional to $\mathrm{Tr}(\hat{x}\ddot{\hat{\rho}}_r)$. With these simplifications,  Eq.~\eqref{eq:AvPTrace} can be written as 
\begin{align}\label{eq:QuantumAbrahamLorentz}
m_{\text{\tiny{R}}}\frac{d^2}{dt^2}\langle\hat{x}\rangle =\mathrm{F}_{\mathrm{ext}} +\frac{2\alpha\hbar}{3c^2}\frac{d^3}{dt^3}\langle\hat{x}\rangle\,,
\end{align}
where $m_{\text{\tiny{R}}}:=m+(4\alpha\hbar\omega_{\text{\tiny{max}}})/(3\pi c^2)$. We notice that Eq.~\eqref{eq:QuantumAbrahamLorentz} has the same form as the Abraham-Lorentz equation~\eqref{eq:AL}. The same result is also obtained for the general case in which the bare  potential explicitly depends on time, such that $\hat{V}_0=\hat{V}_0(x,t)$, as shown in Appendix~\ref{App:AppendixB}. We remark that the equation of motion derived quantum mechanically only reduces to Eq.~\eqref{eq:AL} in the presence of an external potential. When the external potential is switched off, the EOM reduces to Eq.~\eqref{eq:FreeParticle}, for any value of the cutoff $\omega_{\text{\tiny{max}}}$, and is therefore free of the runaway solution.

\section{Decoherence}\label{Sec:Decoherence} In this final part of the article, we are interested in assessing if the spatial superposition of a charged particle at rest can be suppressed via its interaction with the vacuum fluctuations alone. We begin by writing the position space representation of the master equation~\eqref{eq:MasterEquation} relevant for decoherence, 
\begin{align}\label{eq:decoherence}
\partial_t\rho_r=\left[-\frac{(x'-x)^2\mathcal{N}_1(t)}{\hbar}\right]\rho_r\,,
\end{align}
where $\mathcal{N}_1(\tau)$ is defined to be 
\begin{align}
\mathcal{N}_{1}(\tau):=\int_{0}^{\tau} d\tau'{\mathcal{N}(\tau')} = -\frac{4\alpha\hbar}{3\pi c^2}\frac{\tau^3-3\tau \epsilon^2}{(\tau^2+\epsilon^2)^{3}}\,.
\end{align}
We have set $t_i=0$ and only retained the second term involving the noise kernel in Eq.~\eqref{eq:MasterEquation}. This is because the other terms typically give subdominant contributions when the question of interest is to evaluate the rate of decay of the off-diagonal elements of the density matrix at late times \cite{Schlosshauer2007, calzetta_hu_2008}. We have also used the expression of the noise kernel in Eq.~\eqref{eq:NoiseKernel} inside the integral to obtain the expression for $\mathcal{N}_1$. Integrating Eq.~\eqref{eq:decoherence}, we get
\begin{align}\label{eq:Decoherence}
\rho_r(x',x,t)=\exp\left(-\frac{(x'-x)^2}{\hbar}\mathcal{N}_2(t)\right)\rho_r(x',x,0)\,,
\end{align}
where $\mathcal{N}_2(t) := \int_{0}^td\tau\mathcal{N}_1(\tau)$.
The function $\mathcal{N}_2(t)$ is inversely proportional to the coherence length $l_x(t)$ defined by $l_x(t):=\left({\hbar}/{\mathcal{N}_2(t)}\right)^{\frac{1}{2}}$. After performing the integral over $\mathcal{N}_1$, the expression for the coherence length is obtained to be 
\begin{align}\label{eq:Coherence_Length}
l_x(t) =\sqrt{\frac{3\pi c^2}{2\alpha\omega^2_{\text{\tiny{max}}}}\cdot\frac{(t^2+\epsilon^2)^2}{t^4+3t^2\epsilon^2}}\stackrel{t\gg\epsilon}{=} \sqrt{\frac{3\pi}{2\alpha}}\frac{1}{k_{\text{\tiny{max}}}}\,.
\end{align}
We see that the coherence length approaches a constant value on time-scales much larger than $\epsilon=1/\omega_{\text{\tiny{max}}}$ and that its value scales inversely with the UV cutoff. Taken literally, if one sets $k_{\text{\tiny{max}}} = 1/\lambda_{{db}}$, where $\lambda_{{db}}$ is the de Broglie wavelength of the electron, one would arrive at the conclusion that vacuum fluctuations lead to decoherence with the coherence length of the charged particle asymptotically reducing to $l_{x}\approx 25\lambda_{{db}}$ within the time scales $t\approx \lambda_{db}/c$. 

\subsection*{False decoherence}
It is clearly unsatisfactory to have an observable effect explicitly scale with the UV cutoff since the precise numerical value of the cutoff is, strictly speaking, arbitrary. A similar situation was encountered in \cite{Unruh_Coherence} in a different context of a harmonic oscillator coupled to a massive scalar field. However, it was argued in \cite{Unruh_Coherence} that the reduced density matrix of the harmonic oscillator described false decoherence. In such a situation, the off-diagonal elements of the density matrix are suppressed simply because the state of the environment goes into different configurations depending upon the spatial location of the system. However, these changes in the environmental states remain locally around the system and are reversible. For the electron interacting with vacuum fluctuations, we therefore take the point of view that if the reduced density matrix describes false decoherence, then, after adiabatically switching off the interactions with the environment (after having adiabatically switched it on initially), the original coherence must be fully restored at the level of the system.

To formulate the argument, we consider a time-dependent coupling $q(t)=-ef(t)$ such that $f(t)=1$ for most of the dynamics between the initial time $t=0$ and the final time $t=T$, while $f(0)=f(T)=0$. The quantity relevant for decoherence is the noise kernel  which, under the time-dependent coupling, transforms as $\mathcal{N}\rightarrow\tilde{\mathcal{N}}$, with  
\begin{align}
\tilde{\mathcal{N}} = f(t_1)f(t_2)\mathcal{N}(t_1;t_2) =  f(t_1)f(t_2)\mathcal{N}(t_1-t_2)\,.
\end{align}
The decoherence factor in the double commutator in Eq.~\eqref{eq:MasterEquation} involves replacing $t_2$ with $t_1-\tau$ and then integrating over $\tau$. Therefore, the function $\mathcal{N}_1$ transforms as $\mathcal{N}_1\rightarrow\tilde{\mathcal{N}}_1$, with 
\begin{align}\label{eq:TildeN1v1}
\tilde{\mathcal{N}}_{1}(t_1)=f(t_1)\int_{0}^{t_1}d\tau f(t_1-\tau)\mathcal{N}(\tau)\,.
\end{align}
From the definitions of $\mathcal{N}_1$ and $\mathcal{N}_2$, we have $\mathcal{N}= (d/d\tau)\mathcal{N}_1$,  $\mathcal{N}_1= (d/d\tau)\mathcal{N}_2$ and $\mathcal{N}_1(0)=\mathcal{N}_2(0)=0$. Using these relations and integrating by parts, Eq.~\eqref{eq:TildeN1v1} becomes
\begin{align}
\tilde{\mathcal{N}}_{1}(t_1) =&f(t_1)\mathcal{N}_1(t_1)f(0)+f(t_1)\mathcal{N}_2(t_1)\dot{f}(0)\nonumber\\&+f(t_1)\int_{0}^{t_1}d\tau\mathcal{N}_2(\tau)\frac{d^2}{d\tau^2}f(t_1-\tau)\,. 
\end{align}
In the limit $\epsilon\to 0$ (taking the UV cutoff to infinity), we see from Eq.~\eqref{eq:Coherence_Length} that $\mathcal{N}_2$ loses any time dependence. We can therefore bring $\mathcal{N}_2$ outside the integral such that 
\begin{align}
\tilde{\mathcal{N}}_{1}(t_1) =&f(t_1)\mathcal{N}_1(t_1)f(0)+f(t_1)\mathcal{N}_2\dot{f}(0)\nonumber\\
&-f(t_1)\mathcal{N}_2(\dot{f}(0)-\dot{f}(t_1))\,.
\end{align}
The terms involving $\dot{f}(0)$ cancel out and we get
\begin{align}\label{eq:TildeN1}
\tilde{\mathcal{N}}_{1}(t_1)=f(t_1)\mathcal{N}_1(t_1)f(0)+f(t_1)\mathcal{N}_2\dot{f}(t_1)\,.
\end{align}
After integrating by parts in Eq.~\eqref{eq:TildeN1}, in order to obtain $\tilde{\mathcal{N}}_2(T)=\int_{0}^Tdt_1\tilde{\mathcal{N}}_{1}(t_1)$, we get 
\begin{align}
\tilde{\mathcal{N}}_2(T)=&f(0)\left(f(T)\mathcal{N}_2(T)-f(0)\mathcal{N}_2(0)\right)\nonumber\\
&-f(0)\mathcal{N}_2\int_{0}^{T}dt_1\dot{f}+\frac{\mathcal{N}_2}{2}\int_{0}^Tdt_1\frac{d}{dt_1}f^2\,.
\end{align}
In the limit $\epsilon\to0$, as we noted earlier, $\mathcal{N}_2(t)$ takes a constant value for any time $t>0$, but is zero at $t=0$ from the way it is defined. Therefore, after completing the remaining integrals, we get
\begin{align}\label{eq:falsedec}
\tilde{\mathcal{N}}_2(T)=\frac{\mathcal{N}_2}{2}\left(f^2(0)+f^2(T)\right)\,.
\end{align}
Since we assume that the interactions are switched off in the very beginning and at the very end, we see that $\tilde{\mathcal{N}}_2(T)=0$ such that Eq.~\eqref{eq:Decoherence} becomes $\tilde{\rho}_r(x',x,T)=\rho_r(x',x,0)$. 
Therefore, by adiabatically switching off the interactions, we recover the original coherence within the system. 

This is different from standard collisional decoherence where, for example, one originally has $\partial_t{\rho}_r(x',x,t)=-\Lambda (x'-x)^2\rho_r(x',x,t)$ \cite{Schlosshauer2007}. When in this case we send $\Lambda\rightarrow\tilde{\Lambda}=f(t)\Lambda$, we get $\tilde{\rho}_r(x',x,t)=\exp{-\Lambda (x'-x)^2\int_0^t dt'f(t')}\rho_r(x',x,0)$. The density matrix depends on the integral of  $f(t)$ rather than its end points and we see that coherence is indeed lost irreversibly. Thus, we interpret our result~\eqref{eq:falsedec}, which differs from the ones obtained in \cite{Caldeira1991,BandP}, to imply that the vacuum fluctuations alone do not lead to irreversible loss of coherence. 

\section{Discussion}
We formulated the interaction of a nonrelativistic electron with the  radiation field within the framework of open quantum systems and obtained the master equation for the reduced electron dynamics in the position basis. We showed that the classical limit of the quantum dynamics is free of the problems associated with the purely classical derivation of the Abraham-Lorentz equation. With respect to possible decoherence induced by vacuum fluctuations alone, we showed that the apparent decoherence at the level of the reduced density matrix is reversible and is an artifact of the formalism used. In mathematically tracing over the environment, one traces over the degrees of freedom that physically surround the system being observed. These degrees of freedom must be considered part of the system being observed, rather than the environment \cite{Diosi1995,Unruh_Coherence}. We formulated this interpretation by showing that one restores full initial coherence back into the system after switching off the interactions with the environment adiabatically. The formulation is fairly general and might also be used in other situations to distinguish true decoherence from a false one. The analysis therefore brings together various works in the literature \cite{Santos1994,Diosi1995,Caldeira1991,BandP,Unruh_Coherence} and addresses some of the conflicting results.

\section{Acknowledgements}
We thank Davide Bason and Lorenzo Di Pietro for numerous discussions, Oliviero Angeli for cross checking some of the results obtained in the manuscript and Lajos Di\'{o}si for discussions concerning false decoherence. We acknowledge financial support from the  University of Trieste, INFN and the EIC Pathfinder project QuCoM (GA No. 101046973). A.B. also acknowledges financial support from the PNRR MUR project PE0000023-NQSTI.

\appendix

\section{Integrals involving the dissipation kernel}\label{App:AppendixA}
In this appendix, we derive an identity involving the integrals of the form $\int d\tau\mathcal{D}(\tau)f(\tau)$. 
To proceed, we keep in mind the situation where $\epsilon$ is small but finite so that all the derivatives of the \textit{smoothed} Dirac delta $\delta_{\epsilon}(\tau)$ are large but finite. However, for times $\tau\gg \epsilon$, we have $\delta_{\epsilon}(\tau) = \delta_{\epsilon}'(\tau) = \delta_{\epsilon}''(\tau) = 0$. In addition, since the derivative of the Dirac delta is an odd function of $\tau$, we also have $\delta_{\epsilon}'(0) = 0$.  In computing the integral of $\mathcal{D}(\tau)$ multiplying an arbitrary function $f(\tau)$, we shift the derivatives acting on $\delta_{\epsilon}$, one by one, onto $f(\tau)$ by integrating by parts. Since the calculations of interest involve integrating $\int_{0}^td\tau\mathcal{D}(\tau) f(\tau)$, where $\tau$ takes only non-negative values from $0$ to $t$, the Heaviside step function $\theta(\tau)$ can be omitted inside the integral.

The first integration by parts [the constant pre-factors appearing in Eq.~\eqref{eq:DissipationKernel} will be plugged in at the end] gives 
\begin{align}\label{eq:App:thetadef}
\int_{0}^td\tau \delta'''_{\epsilon}(\tau) f(\tau) = 	-\int_{0}^td \tau \delta''_{\epsilon}(\tau)f'(\tau)+\left.\delta''_{\epsilon}(\tau)f(\tau)\right|^{t}_0\,.
\end{align} 
Since $\delta''_{\epsilon}(t)=0$, only the boundary term $-\delta_{\epsilon}''(0)f(0)$ survives. Further,
\begin{align}
-\int_{0}^t d\tau\delta''_{\epsilon}(\tau) f'(\tau) = \int_{0}^t d\tau\delta'_{\epsilon}(\tau)f''(\tau)-\delta'_{\epsilon}(\tau)\left.f'(\tau)\right|^t_{0}\,.
\end{align}
Since $\delta_{\epsilon}'(t) = \delta_{\epsilon}'(0) = 0$ [$\delta_{\epsilon}'(\tau)$ being an odd function of $\tau$], both the boundary terms vanish. Proceeding further, we get
\begin{align}\label{eq:App:divergent_integral}
\int_{0}^t d\tau\delta'_{\epsilon}(\tau)f''(\tau) = -\int_{0}^t d\tau\delta_{\epsilon}(\tau)f'''(\tau) +\delta_{\epsilon}(\tau) \left.f''(\tau)\right|^{t}_{0}\,.
\end{align}
As before, the boundary term at $\tau=t$ is zero and only the term $-\delta_{\epsilon}(0)f''(0)$ survives. Finally, since $\delta_{\epsilon}(\tau)$ goes to zero much faster than a generic function $f(\tau)$ for a small $\epsilon$, it can be treated like a Dirac delta such that 
\begin{align}
-\int_{0}^td\tau\delta_{\epsilon}(\tau)f'''(\tau) = -\frac{f'''(0)}{2}\,.
\end{align}
The factor of half comes because the integral is performed from $0$ to $t$. Collecting the two boundary terms, we get the result
\begin{align}\label{eq:App:integral_final}
\int_{0}^td\tau \delta'''_{\epsilon}(\tau)f(\tau) = -\frac{f'''(0)}{2} - \delta_{\epsilon}(0)f''(0)-\delta''_{\epsilon}(0)f(0)\,. 
\end{align} 
From Eq.~\eqref{eq:deltafunction}, we have $\delta_{\epsilon}(0) = 1/(\pi\epsilon) = \omega_{\text{\tiny{max}}}/\pi$ and $\delta_{\epsilon}''(0)=-2\omega^3_{\text{\tiny{max}}}/\pi$ such that 
\begin{align}\label{eq:App:integral_dissipation_kernel}
\int_{0}^td\tau\mathcal{D}(\tau)f(\tau) = &-\frac{2\alpha\hbar}{3c^2} f'''(0) -\frac{4\alpha\hbar\omega_{\text{\tiny{max}}}}{3\pi c^2} f''(0)\nonumber\\
&+\frac{2e^2\omega^3_{\text{\tiny{max}}}}{3\pi^2\epsilon_0c^3}f(0)\,.
\end{align}
Here, we have now plugged in the constant prefactor appearing in Eq.~\eqref{eq:DissipationKernel}.

\section{The Abraham-Lorentz equation as a classical limit}\label{App:AppendixB}
The rate of change of the expectation values can be obtained with the help of the master equation~\eqref{eq:MasterEquation}. For the position operator, it is given by
\begin{align}\label{Eq:App:EOMX1}
&\frac{d}{dt}\langle\hat{x}\rangle = \mathrm{Tr}\left(\hat{x}\partial_t\hat{\rho}_r\right)=\nonumber\\
&-\frac{i}{\hbar}\mathrm{Tr}\left(\hat{x}\cdot\left[\hat{\mathrm{H}}_s,\hat{\rho}_{r}\right]\right)\nonumber\\ 
&+\frac{i}{2\hbar}\int_{0}^{t-t_i}
d\tau\mathcal{D}(t;t-\tau)\mathrm{Tr}\left(\hat{x}\cdot\left[\hat{x},\{\hat{x}_{\text{\tiny{H}}_s}(-\tau),\hat{\rho}_{r}(t)\}\right]\right)\nonumber\\&-\frac{1}{\hbar}\int_{0}^{t-t_i} d\tau \mathcal{N}(t;t-\tau)\mathrm{Tr}\left(\hat{x}\cdot\left[\hat{x},\left[\hat{x}_{\text{\tiny{H}}_s}(-\tau),\hat{\rho}_{r}(t)\right]\right]\right)\,.
\end{align} 
Due to the identity 
\begin{align}\label{eq:App:CyclicProperty}
\mathrm{Tr}\left(\hat{A}\cdot\left[\hat{B},\hat{C}\right]\right) = \mathrm{Tr}\left(\left[\hat{A},\hat{B}\right]\cdot\hat{C}\right)\,,
\end{align}
the terms involving the dissipation and the noise kernels vanish and we get
\begin{align}\label{eq:App:EOMX}
\frac{d}{dt}\langle\hat{x}\rangle=-\frac{i}{\hbar}\mathrm{Tr}\left(\hat{\rho}_r\cdot\left[\hat{x},\hat{\mathrm{H}}_s\right]\right) = \frac{\langle\hat{p}\rangle}{m}\,.
\end{align}
Here, we remember that the system Hamiltonian $\hat{\mathrm{H}}_s$ receives a contribution from $\hat{V}_{\text{\tiny{EM}}}$ in addition to the bare potential $\hat{V}_0$ such that [cf. the discussion between Eqs.~\eqref{eq:Hs} and~\eqref{eq:VEMExact}]
\begin{align}
\hat{\mathrm{H}}_s(t) = \frac{\hat{p}^2}{2m}+\hat{V}_0(x,t)+\frac{e^2\omega^3_{\text{\tiny{max}}}}{3\pi^2\epsilon_0c^3}\hat{x}^2\,.
\end{align}
Proceeding analogously, we obtain, for the momentum operator, 
\begin{align}\label{eq:App:EOMP1}
&\frac{d}{dt}\langle\hat{p}\rangle = \mathrm{Tr}\left(\hat{p}\partial_t\hat{\rho}_r\right)=\nonumber\\
&-\frac{i}{\hbar}\mathrm{Tr}\left(\left[\hat{p},\hat{\mathrm{H}}_s\right]\cdot\hat{\rho}_{r}\right)\nonumber\\ 
&+\frac{i}{2\hbar}\int_{0}^{t-t_i}
d\tau\mathcal{D}(t;t-\tau)\mathrm{Tr}\left(\left[\hat{p},\hat{x}\right]\cdot\{\hat{x}_{\text{\tiny{H}}_s}(-\tau),\hat{\rho}_{r}(t)\}\right)\nonumber\\&-\frac{1}{\hbar}\int_{0}^{t-t_i} d\tau \mathcal{N}(t;t-\tau)\mathrm{Tr}\left(\left[\hat{p},\hat{x}\right]\cdot\left[\hat{x}_{\text{\tiny{H}}_s}(-\tau),\hat{\rho}_{r}(t)\right]\right)\,.
\end{align} 
Since $\left[\hat{x},\hat{p}\right] = i\hbar\mathds{1}$, the term involving the noise kernel vanishes and Eq.~\eqref{eq:App:EOMP1} simplifies to
\begin{align}\label{eq:App:AvP1}
\frac{d}{dt}\langle\hat{p}\rangle = &-\langle\hat{V}_0,_{{x}}\rangle -\frac{2e^2\omega^3_{\text{\tiny{max}}}}{3\pi^2\epsilon_0c^3}\langle\hat{x}\rangle\nonumber\\
&+\mathrm{Tr}\left(\hat{\rho}_{r}(t)\int_{0}^{t-t_i}d\tau \mathcal{D}(\tau)\hat{x}_{\text{\tiny{H}}_s}(-\tau)\right)\,.
\end{align}
Evaluating the integral using Eq.~\eqref{eq:App:integral_dissipation_kernel}, we see that the last term in the second line of Eq.~\eqref{eq:App:integral_dissipation_kernel} gives the contribution $\frac{2e^2\omega^3_{\text{\tiny{max}}}}{3\pi^2\epsilon_0c^3}\langle\hat{x}\rangle$ to $\frac{d}{dt}\langle\hat{p}\rangle$ in Eq.~\eqref{eq:App:AvP1} and  cancels the contribution coming from $\hat{V}_{\text{\tiny{EM}}}$. The EOM therefore reduces to
\begin{align}\label{eq:App:AvP2}
\frac{d}{dt}\langle\hat{p}\rangle = &-\langle\hat{V_0},_{{x}}\rangle-\frac{4\alpha\hbar\omega_{\text{\tiny{max}}}}{3\pi c^2}\mathrm{Tr}\left(\hat{\rho}_{r}(t)\left.\frac{d^2}{d\tau^2}\hat{x}_{\text{\tiny{H}}_s}(-\tau)\right|_{\tau=0}\right)\nonumber\\
&-\frac{2\alpha\hbar}{3c^2}\mathrm{Tr}\left(\hat{\rho}_{r}(t)\left.\frac{d^3}{d\tau^3}\hat{x}_{\text{\tiny{H}}_s}(-\tau)\right|_{\tau=0}\right)\,.
\end{align}
As shown in the main article, when $\hat{V}_0=0$, the double and the triple derivatives acting on $\hat{x}_{\text{\tiny{H}}_s}(-\tau)$ vanish up to second order in the interactions in Eq.~\eqref{eq:App:AvP2}, thereby making it free of the instability problems associated with the classical AL equation. Here, we only focus on the general case in which the external (time-dependent) potential is switched on. To  simplify the equation further, we begin by evaluating the second-order derivative in Eq.~\eqref{eq:App:AvP2}. From Eq.~\eqref{eq:xH}, we have
\begin{align}\label{eq:App:xHDoubleDerivative}
\frac{d^2}{d\tau^2}\hat{x}_{\text{\tiny{H}}_s}(-\tau) ={}&  \hat{U}_s^{-1}(t-\tau;t)\hat{x}\hat{U}_s''(t-\tau;t)\nonumber\\
&+2\hat{U}_s^{-1'}(t-\tau;t)\hat{x}\hat{U}_s'(t-\tau;t)\nonumber\\
&+\hat{U}_s^{-1''}(t-\tau;t)\hat{x}\hat{U}_s(t-\tau;t)\,,
\end{align}
where the prime denotes the derivative with respect to $\tau$. From the Schr\"{o}dinger equation 
\begin{align}
\hat{U}_s'(t-\tau;t) = \frac{i}{\hbar}\hat{\mathrm{H}}_s(t-\tau)\hat{U}_s(t-\tau;t)\,,
\end{align}
the derivatives acting on the unitary operator can be expressed in terms of the Hamiltonian. It is clear that taking higher derivatives of $\hat{U}_s(t-\tau;t)$ would result in  higher powers of the Hamiltonian or the partial derivative of the Hamiltonian with respect to $\tau$, multiplied with only a single unitary operator on the very right. However, if in the end $\tau$ is set to zero, the Hamiltonian and its explicit time derivatives will be evaluated at time $t$, and the unitary operator on the very right disappears since $\hat{U}_s(t;t) = \mathds{1}$. We therefore have the following identities:
\begin{align}
\left.\hat{U}_s^{(\prime n)}(t-\tau;t)\right|_{\tau=0} = (-1)^n\left(\frac{d^n}{dt^n}\hat{U}_s (t;t_i)\right)\hat{U}_s^{-1}(t;t_i)\,,\label{eq:App:UDerivative}\\
\left.\hat{U}_s^{-1(\prime n)}(t-\tau;t)\right|_{\tau=0} = (-1)^n\hat{U}_s(t;t_i)\left(\frac{d^n}{dt^n}\hat{U}_s^{-1} (t;t_i)\right)\,.\label{eq:App:UIDerivative}
\end{align}
The additional time parameter $t_i$ that appears in Eqs.~\eqref{eq:App:UDerivative} and~\eqref{eq:App:UIDerivative} is only apparent. As discussed before, evaluating the time derivatives on the right-hand side of Eq.~\eqref{eq:App:UDerivative} would result in powers of $\hat{\mathrm{H}}_s(t)$ and its derivatives evaluated at $t$. The remaining unitary matrix $\hat{U}_s(t;t_i)$ would be canceled by the additional $\hat{U}_s^{-1}(t;t_i)$ on the very right such that $t_i$ disappears from the equation.
Using Eqs.~\eqref{eq:App:UDerivative} and~\eqref{eq:App:UIDerivative} in Eq.~\eqref{eq:App:xHDoubleDerivative}, we get
\begin{align}\label{eq:App:Trace_DoubleDerivative_Long}
&\mathrm{Tr}\left(\hat{\rho}_{r}(t)\left.\frac{d^2}{d\tau^2}\hat{x}_{\text{\tiny{H}}_s}(-\tau)\right|_{\tau=0}\right) =\nonumber\\ &\mathrm{Tr}\left(\left[\left(\frac{d^2}{dt^2}\hat{U}_s (t;t_i)\right)\hat{U}_s^{-1}(t;t_i)\hat{\rho}_{r}(t)+\right.\right.\nonumber\\
&2\left(-\frac{d}{dt}\hat{U}_s (t;t_i)\right)\hat{U}_s^{-1}(t;t_i)\hat{\rho}_{r}(t)\hat{U}_s(t;t_i)\left(-\frac{d}{dt}\hat{U}_s^{-1} (t;t_i)\right)\nonumber\\&\left.\left.+\hat{\rho}_{r}(t)\hat{U}_s(t;t_i)\left(\frac{d^2}{dt^2}\hat{U}_s^{-1} (t;t_i)\right)\right]\hat{x}\right)\,.
\end{align}
Here, we have used the cyclic property within the trace to shift the unitary operator $\hat{U}_s$ and its derivatives on the right of $\hat{x}$ in Eq.~\eqref{eq:App:xHDoubleDerivative} onto the very left within the trace.

To proceed further, we note that the terms involving the trace in Eq.~\eqref{eq:App:AvP2} are multiplied by $\alpha$. It is therefore sufficient to evaluate the trace at zeroth order in the interactions as the master equation is valid only up to second order in the interactions. This implies that within the trace, the time dependence of the density matrix can be evaluated  by keeping only the Liouville--von Neumann term such that
\begin{align}\label{eq:App:RhoLV}
\hat{\rho}_r(t) = \hat{U}_s(t;t_i)\hat{\rho}_r(t_i)\hat{U}^{-1}_s(t;t_i)\,.
\end{align}
Equation~\eqref{eq:App:Trace_DoubleDerivative_Long} then simplifies to
\begin{align}\label{eq:App:DDposition}
&\mathrm{Tr}\left(\hat{\rho}_{r}(t)\left.\frac{d^2}{d\tau^2}\hat{x}_{\text{\tiny{H}}_s}(-\tau)\right|_{\tau=0}\right) =\nonumber\\ &\quad\mathrm{Tr}\left(\left[\left(\frac{d^2}{dt^2}\hat{U}_s (t;t_i)\right)\hat{\rho}_{r}(t_i)\hat{U}_s^{-1}(t;t_i)\right.\right.\nonumber\\
&\qquad\qquad+2\left(\frac{d}{dt}\hat{U}_s (t;t_i)\right)\hat{\rho}_{r}(t_i)\left(\frac{d}{dt}\hat{U}_s^{-1} (t;t_i)\right)\nonumber\\
&\qquad\qquad\left.\left.+\hat{U}_s(t;t_i)\hat{\rho}_{r}(t_i)\left(\frac{d^2}{dt^2}\hat{U}_s^{-1} (t;t_i)\right)\right]\hat{x}\right)\,.
\end{align}
Thus, from Eqs.~\eqref{eq:App:RhoLV} and~\eqref{eq:App:DDposition}, we get
\begin{align}\label{eq:App:TraceDouble}
\mathrm{Tr}\left(\hat{\rho}_{r}(t)\left.\frac{d^2}{d\tau^2}\hat{x}_{\text{\tiny{H}}_s}(-\tau)\right|_{\tau=0}\right)&={}\mathrm{Tr}\left(\ddot{\hat{\rho}}_r(t)\hat{x}\right)=\frac{d^2}{dt^2}\langle\hat{x}\rangle\,.
\end{align}
A similar line of reasoning also leads to the relation
\begin{align}\label{eq:App:TraceTriple}
\mathrm{Tr}\left(\hat{\rho}_{r}(t)\left.\frac{d^3}{d\tau^3}\hat{x}_{\text{\tiny{H}}_s}(-\tau)\right|_{\tau=0}\right)=-\frac{d^3}{dt^3}\langle\hat{x}\rangle\,.
\end{align}
Using Eqs.~\eqref{eq:App:TraceDouble} and~\eqref{eq:App:TraceTriple} in Eq.~\eqref{eq:App:AvP2}, the EOM for the expectation value of the position operator in the presence of an external potential is obtained to be
\begin{align}\label{eq:App:QuantumAbrahamLorentz}
m_{\text{\tiny{R}}}\frac{d^2}{dt^2}\langle\hat{x}\rangle &=-\langle\hat{V}_0,_{x}\rangle +\frac{2\alpha\hbar}{3c^2}\frac{d^3}{dt^3}\langle\hat{x}\rangle\,,
\end{align}
where $m_{\text{\tiny{R}}}:=m+\frac{4\alpha\hbar\omega_{\text{\tiny{max}}}}{3\pi c^2}$.

\begin{thebibliography}{33}%
	\makeatletter
	\providecommand \@ifxundefined [1]{%
		\@ifx{#1\undefined}
	}%
	\providecommand \@ifnum [1]{%
		\ifnum #1\expandafter \@firstoftwo
		\else \expandafter \@secondoftwo
		\fi
	}%
	\providecommand \@ifx [1]{%
		\ifx #1\expandafter \@firstoftwo
		\else \expandafter \@secondoftwo
		\fi
	}%
	\providecommand \natexlab [1]{#1}%
	\providecommand \enquote  [1]{``#1''}%
	\providecommand \bibnamefont  [1]{#1}%
	\providecommand \bibfnamefont [1]{#1}%
	\providecommand \citenamefont [1]{#1}%
	\providecommand \href@noop [0]{\@secondoftwo}%
	\providecommand \href [0]{\begingroup \@sanitize@url \@href}%
	\providecommand \@href[1]{\@@startlink{#1}\@@href}%
	\providecommand \@@href[1]{\endgroup#1\@@endlink}%
	\providecommand \@sanitize@url [0]{\catcode `\\12\catcode `\$12\catcode
		`\&12\catcode `\#12\catcode `\^12\catcode `\_12\catcode `\%12\relax}%
	\providecommand \@@startlink[1]{}%
	\providecommand \@@endlink[0]{}%
	\providecommand \url  [0]{\begingroup\@sanitize@url \@url }%
	\providecommand \@url [1]{\endgroup\@href {#1}{\urlprefix }}%
	\providecommand \urlprefix  [0]{URL }%
	\providecommand \Eprint [0]{\href }%
	\providecommand \doibase [0]{https://doi.org/}%
	\providecommand \selectlanguage [0]{\@gobble}%
	\providecommand \bibinfo  [0]{\@secondoftwo}%
	\providecommand \bibfield  [0]{\@secondoftwo}%
	\providecommand \translation [1]{[#1]}%
	\providecommand \BibitemOpen [0]{}%
	\providecommand \bibitemStop [0]{}%
	\providecommand \bibitemNoStop [0]{.\EOS\space}%
	\providecommand \EOS [0]{\spacefactor3000\relax}%
	\providecommand \BibitemShut  [1]{\csname bibitem#1\endcsname}%
	\let\auto@bib@innerbib\@empty
	\bibitem [{\citenamefont {Casimir}(1948)}]{Casimir1948}%
	\BibitemOpen
	\bibfield  {author} {\bibinfo {author} {\bibfnamefont {H.~B.~G.}\
			\bibnamefont {Casimir}},\ }\href@noop {} {\bibfield  {journal} {\bibinfo
			{journal} {Indag. Math.}\ }\textbf {\bibinfo {volume} {10}},\ \bibinfo
		{pages} {261} (\bibinfo {year} {1948})}\BibitemShut {NoStop}%
	\bibitem [{\citenamefont {Birrell}\ and\ \citenamefont
		{Davies}(1984)}]{Birrell}%
	\BibitemOpen
	\bibfield  {author} {\bibinfo {author} {\bibfnamefont {N.~D.}\ \bibnamefont
			{Birrell}}\ and\ \bibinfo {author} {\bibfnamefont {P.~C.~W.}\ \bibnamefont
			{Davies}},\ }\href {https://doi.org/10.1017/CBO9780511622632} {\emph
		{\bibinfo {title} {{Quantum Fields in Curved Space}}}},\ Cambridge Monographs
	on Mathematical Physics\ (\bibinfo  {publisher} {Cambridge University Press},\
	\bibinfo {address} {Cambridge, UK},\ \bibinfo {year} {1984})\BibitemShut
	{NoStop}%
	\bibitem [{\citenamefont {Parker}\ and\ \citenamefont
		{Toms}(2009)}]{parker_toms_2009}%
	\BibitemOpen
	\bibfield  {author} {\bibinfo {author} {\bibfnamefont {L.}~\bibnamefont
			{Parker}}\ and\ \bibinfo {author} {\bibfnamefont {D.}~\bibnamefont {Toms}},\
	}\href {https://doi.org/10.1017/CBO9780511813924} {\emph {\bibinfo {title}
			{Quantum Field Theory in Curved Spacetime: Quantized Fields and Gravity}}},\
	Cambridge Monographs on Mathematical Physics\ (\bibinfo  {publisher}
	{Cambridge University Press},\
	\bibinfo {address} {Cambridge, UK},\ \bibinfo {year} {2009})\BibitemShut {NoStop}%
	\bibitem [{\citenamefont {Unruh}(1976)}]{UnruhEffect}%
	\BibitemOpen
	\bibfield  {author} {\bibinfo {author} {\bibfnamefont {W.~G.}\ \bibnamefont
			{Unruh}},\ }\href {https://doi.org/10.1103/PhysRevD.14.870} {\bibfield
		{journal} {\bibinfo  {journal} {Phys. Rev. D}\ }\textbf {\bibinfo {volume}
			{14}},\ \bibinfo {pages} {870} (\bibinfo {year} {1976})}\BibitemShut
	{NoStop}%
	\bibitem [{\citenamefont {Fulling}(1973)}]{Fulling}%
	\BibitemOpen
	\bibfield  {author} {\bibinfo {author} {\bibfnamefont {S.~A.}\ \bibnamefont
			{Fulling}},\ }\href {https://doi.org/10.1103/PhysRevD.7.2850} {\bibfield
		{journal} {\bibinfo  {journal} {Phys. Rev. D}\ }\textbf {\bibinfo {volume}
			{7}},\ \bibinfo {pages} {2850} (\bibinfo {year} {1973})}\BibitemShut
	{NoStop}%
	\bibitem [{\citenamefont {Takagi}(1986)}]{Takagi}%
	\BibitemOpen
	\bibfield  {author} {\bibinfo {author} {\bibfnamefont {S.}~\bibnamefont
			{Takagi}},\ }\href {https://doi.org/10.1143/PTP.88.1} {\bibfield  {journal}
		{\bibinfo  {journal} {Prog. Theor. Phys. Suppl.}\ }\textbf {\bibinfo {volume}
			{88}},\ \bibinfo {pages} {1} (\bibinfo {year} {1986})}\BibitemShut {NoStop}%
	\bibitem [{\citenamefont {Bethe}(1947)}]{BetheLambShift}%
	\BibitemOpen
	\bibfield  {author} {\bibinfo {author} {\bibfnamefont {H.~A.}\ \bibnamefont
			{Bethe}},\ }\href {https://doi.org/10.1103/PhysRev.72.339} {\bibfield
		{journal} {\bibinfo  {journal} {Phys. Rev.}\ }\textbf {\bibinfo {volume}
			{72}},\ \bibinfo {pages} {339} (\bibinfo {year} {1947})}\BibitemShut
	{NoStop}%
	\bibitem [{\citenamefont {Lamb}\ and\ \citenamefont
		{Retherford}(1947)}]{LambLambShift}%
	\BibitemOpen
	\bibfield  {author} {\bibinfo {author} {\bibfnamefont {W.~E.}\ \bibnamefont
			{Lamb}}\ and\ \bibinfo {author} {\bibfnamefont {R.~C.}\ \bibnamefont
			{Retherford}},\ }\href {https://doi.org/10.1103/PhysRev.72.241} {\bibfield
		{journal} {\bibinfo  {journal} {Phys. Rev.}\ }\textbf {\bibinfo {volume}
			{72}},\ \bibinfo {pages} {241} (\bibinfo {year} {1947})}\BibitemShut
	{NoStop}%
	\bibitem [{\citenamefont {Welton}(1948)}]{Welton}%
	\BibitemOpen
	\bibfield  {author} {\bibinfo {author} {\bibfnamefont {T.~A.}\ \bibnamefont
			{Welton}},\ }\href {https://doi.org/10.1103/PhysRev.74.1157} {\bibfield
		{journal} {\bibinfo  {journal} {Phys. Rev.}\ }\textbf {\bibinfo {volume}
			{74}},\ \bibinfo {pages} {1157} (\bibinfo {year} {1948})}\BibitemShut
	{NoStop}%
	\bibitem [{\citenamefont {Dalibard}\ \emph {et~al.}(1982)\citenamefont
		{Dalibard}, \citenamefont {Dupont-Roc},\ and\ \citenamefont
		{Cohen-Tannoudji}}]{dalibard}%
	\BibitemOpen
	\bibfield  {author} {\bibinfo {author} {\bibfnamefont {J.}~\bibnamefont
			{Dalibard}}, \bibinfo {author} {\bibfnamefont {J.}~\bibnamefont
			{Dupont-Roc}},\ and\ \bibinfo {author} {\bibfnamefont {C.}~\bibnamefont
			{Cohen-Tannoudji}},\ }\href
	{https://doi.org/10.1051/jphys:0198200430110161700} {\bibfield  {journal}
		{\bibinfo  {journal} {{Journal de Physique}}\ }\textbf {\bibinfo {volume}
			{43}},\ \bibinfo {pages} {1617} (\bibinfo {year} {1982})}\BibitemShut
	{NoStop}%
	\bibitem [{\citenamefont {Joos}\ \emph {et~al.}(2003)\citenamefont {Joos},
		\citenamefont {Zeh}, \citenamefont {Giulini}, \citenamefont {Kiefer},
		\citenamefont {Kupsch},\ and\ \citenamefont
		{Stamatescu}}]{KieferDecoherence}%
	\BibitemOpen
	\bibfield  {author} {\bibinfo {author} {\bibfnamefont {E.}~\bibnamefont
			{Joos}}, \bibinfo {author} {\bibfnamefont {H.}~\bibnamefont {Zeh}}, \bibinfo
		{author} {\bibfnamefont {D.}~\bibnamefont {Giulini}}, \bibinfo {author}
		{\bibfnamefont {C.}~\bibnamefont {Kiefer}}, \bibinfo {author} {\bibfnamefont
			{J.}~\bibnamefont {Kupsch}},\ and\ \bibinfo {author} {\bibfnamefont
			{I.}~\bibnamefont {Stamatescu}},\ }\href
	{https://books.google.it/books?id=6eTHcxeNxdUC} {\emph {\bibinfo {title}
			{Decoherence and the Appearance of a Classical World in Quantum Theory}}},\
	Physics and astronomy online library\ (\bibinfo  {publisher} {Springer},\
	\bibinfo {address} {New York},\
	\bibinfo {year} {2003})\BibitemShut {NoStop}%
	\bibitem [{\citenamefont {Kiefer}(1992)}]{Kiefer:1992}%
	\BibitemOpen
	\bibfield  {author} {\bibinfo {author} {\bibfnamefont {C.}~\bibnamefont
			{Kiefer}},\ }\href {https://doi.org/10.1103/PhysRevD.46.1658} {\bibfield
		{journal} {\bibinfo  {journal} {Phys. Rev. D}\ }\textbf {\bibinfo {volume}
			{46}},\ \bibinfo {pages} {1658} (\bibinfo {year} {1992})}\BibitemShut
	{NoStop}%
	\bibitem [{\citenamefont {Ford}(1993)}]{Ford}%
	\BibitemOpen
	\bibfield  {author} {\bibinfo {author} {\bibfnamefont {L.~H.}\ \bibnamefont
			{Ford}},\ }\href {https://doi.org/10.1103/PhysRevD.47.5571} {\bibfield
		{journal} {\bibinfo  {journal} {Phys. Rev. D}\ }\textbf {\bibinfo {volume}
			{47}},\ \bibinfo {pages} {5571} (\bibinfo {year} {1993})}\BibitemShut
	{NoStop}%
	\bibitem [{\citenamefont {Baym}\ and\ \citenamefont
		{Ozawa}(2009)}]{Baym_Ozawa}%
	\BibitemOpen
	\bibfield  {author} {\bibinfo {author} {\bibfnamefont {G.}~\bibnamefont
			{Baym}}\ and\ \bibinfo {author} {\bibfnamefont {T.}~\bibnamefont {Ozawa}},\
	}\href {https://doi.org/10.1073/pnas.0813239106} {\bibfield  {journal}
		{\bibinfo  {journal} {Proceedings of the National Academy of Sciences}\
		}\textbf {\bibinfo {volume} {106}},\ \bibinfo {pages} {3035} (\bibinfo {year}
		{2009})}\BibitemShut {NoStop}%
	\bibitem [{\citenamefont {Santos}(1994)}]{Santos1994}%
	\BibitemOpen
	\bibfield  {author} {\bibinfo {author} {\bibfnamefont {E.}~\bibnamefont
			{Santos}},\ }\href
	{https://doi.org/https://doi.org/10.1016/0375-9601(94)90438-3} {\bibfield
		{journal} {\bibinfo  {journal} {Physics Letters A}\ }\textbf {\bibinfo
			{volume} {188}},\ \bibinfo {pages} {198} (\bibinfo {year}
		{1994})}\BibitemShut {NoStop}%
	\bibitem [{\citenamefont {Di\'{o}si}(1995)}]{Diosi1995}%
	\BibitemOpen
	\bibfield  {author} {\bibinfo {author} {\bibfnamefont {L.}~\bibnamefont
			{Di\'{o}si}},\ }\href
	{https://doi.org/https://doi.org/10.1016/0375-9601(94)00846-H} {\bibfield
		{journal} {\bibinfo  {journal} {Physics Letters A}\ }\textbf {\bibinfo
			{volume} {197}},\ \bibinfo {pages} {183} (\bibinfo {year}
		{1995})}\BibitemShut {NoStop}%
	\bibitem [{\citenamefont {Barone}\ and\ \citenamefont
		{Caldeira}(1991)}]{Caldeira1991}%
	\BibitemOpen
	\bibfield  {author} {\bibinfo {author} {\bibfnamefont {P.~M. V.~B.}\
			\bibnamefont {Barone}}\ and\ \bibinfo {author} {\bibfnamefont {A.~O.}\
			\bibnamefont {Caldeira}},\ }\href {https://doi.org/10.1103/PhysRevA.43.57}
	{\bibfield  {journal} {\bibinfo  {journal} {Phys. Rev. A}\ }\textbf {\bibinfo
			{volume} {43}},\ \bibinfo {pages} {57} (\bibinfo {year} {1991})}\BibitemShut
	{NoStop}%
	\bibitem [{\citenamefont {Breuer}\ and\ \citenamefont
		{Petruccione}(2000)}]{BandP}%
	\BibitemOpen
	\bibfield  {author} {\bibinfo {author} {\bibfnamefont {H.-P.}\ \bibnamefont
			{Breuer}}\ and\ \bibinfo {author} {\bibfnamefont {F.}~\bibnamefont
			{Petruccione}},\ }in\ \href
	{https://link.springer.com/chapter/10.1007/3-540-45369-5_3} {\emph {\bibinfo
			{booktitle} {Relativistic Quantum Measurement and Decoherence}}},\ \bibinfo
	{editor} {edited by\ \bibinfo {editor} {\bibfnamefont {H.-P.}\ \bibnamefont
			{Breuer}}\ and\ \bibinfo {editor} {\bibfnamefont {F.}~\bibnamefont
			{Petruccione}}}\ (\bibinfo  {publisher} {Springer Berlin Heidelberg},\
	\bibinfo {address} {Berlin, Heidelberg},\ \bibinfo {year} {2000})\ pp.\
	\bibinfo {pages} {31--65}\BibitemShut {NoStop}%
	\bibitem [{\citenamefont {Coleman}(1982)}]{Coleman1982}%
	\BibitemOpen
	\bibfield  {author} {\bibinfo {author} {\bibfnamefont {S.}~\bibnamefont
			{Coleman}},\ }\bibinfo {title} {Classical electron theory from a modern
		standpoint},\ in\ \href {https://doi.org/10.1007/978-1-4757-0650-5_6} {\emph
		{\bibinfo {booktitle} {Electromagnetism: Paths to Research}}},\ \bibinfo
	{editor} {edited by\ \bibinfo {editor} {\bibfnamefont {D.}~\bibnamefont
			{Teplitz}}}\ (\bibinfo  {publisher} {Springer US},\ \bibinfo {address}
	{Boston, MA},\ \bibinfo {year} {1982})\ pp.\ \bibinfo {pages}
	{183--210}\BibitemShut {NoStop}%
	\bibitem [{\citenamefont {Pearle}(1982)}]{Pearle1982}%
	\BibitemOpen
	\bibfield  {author} {\bibinfo {author} {\bibfnamefont {P.}~\bibnamefont
			{Pearle}},\ }\bibinfo {title} {Classical electron models},\ in\ \href
	{https://link.springer.com/chapter/10.1007/978-1-4757-0650-5_7} {\emph
		{\bibinfo {booktitle} {Electromagnetism: Paths to Research}}},\ \bibinfo
	{editor} {edited by\ \bibinfo {editor} {\bibfnamefont {D.}~\bibnamefont
			{Teplitz}}}\ (\bibinfo  {publisher} {Springer US},\ \bibinfo {address}
	{Boston, MA},\ \bibinfo {year} {1982})\ pp.\ \bibinfo {pages}
	{211--295}\BibitemShut {NoStop}%
	\bibitem [{\citenamefont {Griffiths}(2017)}]{Griffiths2017}%
	\BibitemOpen
	\bibfield  {author} {\bibinfo {author} {\bibfnamefont {D.~J.}\ \bibnamefont
			{Griffiths}},\ }\href {https://doi.org/10.1017/9781108333511} {\emph
		{\bibinfo {title} {Introduction to Electrodynamics}}},\ \bibinfo {edition}
	{4th}\ ed.\ (\bibinfo  {publisher} {Cambridge University Press},\
	\bibinfo {address} {Cambridge, UK},\ \bibinfo
	{year} {2017})\BibitemShut {NoStop}%
	\bibitem [{\citenamefont {Cohen-Tannoudji}\ \emph
		{et~al.}(1997{\natexlab{a}})\citenamefont {Cohen-Tannoudji}, \citenamefont
		{Dupont-Roc},\ and\ \citenamefont {Grynberg}}]{TannoudjiPartOneChapterOne}%
	\BibitemOpen
	\bibfield  {author} {\bibinfo {author} {\bibfnamefont {C.}~\bibnamefont
			{Cohen-Tannoudji}}, \bibinfo {author} {\bibfnamefont {J.}~\bibnamefont
			{Dupont-Roc}},\ and\ \bibinfo {author} {\bibfnamefont {G.}~\bibnamefont
			{Grynberg}},\ }\bibinfo {title} {Classical electrodynamics: The fundamental
		equations and the dynamical variables},\ in\ \href
	{https://doi.org/https://doi.org/10.1002/9783527618422.ch1} {\emph {\bibinfo
			{booktitle} {Photons and Atoms}}}\ (\bibinfo  {publisher} {Wiley},\
	\bibinfo {address} {New York},\ \bibinfo {year} {1997}),\ Chap.~\bibinfo {chapter} {1}, pp.\
	\bibinfo {pages} {5--77}\BibitemShut {NoStop}%
	\bibitem [{\citenamefont {Johnson}\ and\ \citenamefont {Hu}(2002)}]{Hu2002}%
	\BibitemOpen
	\bibfield  {author} {\bibinfo {author} {\bibfnamefont {P.~R.}\ \bibnamefont
			{Johnson}}\ and\ \bibinfo {author} {\bibfnamefont {B.~L.}\ \bibnamefont
			{Hu}},\ }\href {https://doi.org/10.1103/PhysRevD.65.065015} {\bibfield
		{journal} {\bibinfo  {journal} {Phys. Rev. D}\ }\textbf {\bibinfo {volume}
			{65}},\ \bibinfo {pages} {065015} (\bibinfo {year} {2002})}\BibitemShut
	{NoStop}%
	\bibitem [{\citenamefont {Calzetta}\ and\ \citenamefont
		{Hu}(2008)}]{calzetta_hu_2008}%
	\BibitemOpen
	\bibfield  {author} {\bibinfo {author} {\bibfnamefont {E.~A.}\ \bibnamefont
			{Calzetta}}\ and\ \bibinfo {author} {\bibfnamefont {B.-L.~B.}\ \bibnamefont
			{Hu}},\ }\href {https://doi.org/10.1017/CBO9780511535123} {\emph {\bibinfo
			{title} {Nonequilibrium Quantum Field Theory}}},\ Cambridge Monographs on
	Mathematical Physics\ (\bibinfo  {publisher} {Cambridge University Press},\
	\bibinfo {address} {Cambridge, UK},\
	\bibinfo {year} {2008})\BibitemShut {NoStop}%
	\bibitem [{Tan(1997)}]{TannoudjiPartOneChapterTwo}%
	\BibitemOpen
	\bibinfo {title} {Lagrangian and Hamiltonian approach to electrodynamics, the
		standard Lagrangian and the Coulomb gauge},\ in\ \href
	{https://doi.org/https://doi.org/10.1002/9783527618422.ch2} {\emph {\bibinfo
			{booktitle} {Photons and Atoms}}}\ (\bibinfo  {publisher} {Wiley},\
	\bibinfo {address} {New York},\ \bibinfo {year} {1997}),\ Chap.~\bibinfo {chapter} {2}, pp.\
	\bibinfo {pages} {79--168}\BibitemShut {NoStop}%
	\bibitem [{\citenamefont {Altland}\ and\ \citenamefont
		{Simons}(2010)}]{altland_simons_2010}%
	\BibitemOpen
	\bibfield  {author} {\bibinfo {author} {\bibfnamefont {A.}~\bibnamefont
			{Altland}}\ and\ \bibinfo {author} {\bibfnamefont {B.~D.}\ \bibnamefont
			{Simons}},\ }\href {https://doi.org/10.1017/CBO9780511789984} {\emph
		{\bibinfo {title} {Condensed Matter Field Theory}}},\ \bibinfo {edition}
	{2nd}\ ed.\ (\bibinfo  {publisher} {Cambridge University Press},\
	\bibinfo {address} {Cambridge, UK},\ \bibinfo
	{year} {2010})\BibitemShut {NoStop}%
	\bibitem [{Note1()}]{Note1}%
	\BibitemOpen
	\bibinfo {note} {Note that the precise choice of the basis states is
		unimportant since the reduced density matrix is obtained after tracing over
		the environment}\BibitemShut {NoStop}%
	\bibitem [{\citenamefont {Feynman}\ and\ \citenamefont
		{Vernon}(1963)}]{Feynman_Vernon}%
	\BibitemOpen
	\bibfield  {author} {\bibinfo {author} {\bibfnamefont {R.}~\bibnamefont
			{Feynman}}\ and\ \bibinfo {author} {\bibfnamefont {F.}~\bibnamefont
			{Vernon}},\ }\href
	{https://doi.org/https://doi.org/10.1016/0003-4916(63)90068-X} {\bibfield
		{journal} {\bibinfo  {journal} {Annals of Physics}\ }\textbf {\bibinfo
			{volume} {24}},\ \bibinfo {pages} {118} (\bibinfo {year} {1963})}\BibitemShut
	{NoStop}%
	\bibitem [{\citenamefont {Cohen-Tannoudji}\ \emph
		{et~al.}(1997{\natexlab{b}})\citenamefont {Cohen-Tannoudji}, \citenamefont
		{Dupont-Roc},\ and\ \citenamefont {Grynberg}}]{TannoudjiPartOneChapterThree}%
	\BibitemOpen
	\bibfield  {author} {\bibinfo {author} {\bibfnamefont {C.}~\bibnamefont
			{Cohen-Tannoudji}}, \bibinfo {author} {\bibfnamefont {J.}~\bibnamefont
			{Dupont-Roc}},\ and\ \bibinfo {author} {\bibfnamefont {G.}~\bibnamefont
			{Grynberg}},\ }\bibinfo {title} {Quantum electrodynamics in the Coulomb
		gauge},\ in\ \href
	{https://doi.org/https://doi.org/10.1002/9783527618422.ch3} {\emph {\bibinfo
			{booktitle} {Photons and Atoms}}}\ (\bibinfo  {publisher} {Wiley},\
	\bibinfo {address} {New York},\ \bibinfo {year} {1997}),\ Chap.~\bibinfo {chapter} {3}, pp.\
	\bibinfo {pages} {169--252}\BibitemShut {NoStop}%
	\bibitem [{\citenamefont {Griffiths}\ \emph {et~al.}(2010)\citenamefont
		{Griffiths}, \citenamefont {Proctor},\ and\ \citenamefont
		{Schroeter}}]{Griffiths2010}%
	\BibitemOpen
	\bibfield  {author} {\bibinfo {author} {\bibfnamefont {D.~J.}\ \bibnamefont
			{Griffiths}}, \bibinfo {author} {\bibfnamefont {T.~C.}\ \bibnamefont
			{Proctor}},\ and\ \bibinfo {author} {\bibfnamefont {D.~F.}\ \bibnamefont
			{Schroeter}},\ }\href {https://aapt.scitation.org/doi/10.1119/1.3269900}
	{\bibfield  {journal} {\bibinfo  {journal} {American Journal of Physics}\
		}\textbf {\bibinfo {volume} {78}},\ \bibinfo {pages} {391} (\bibinfo {year}
		{2010})}\BibitemShut {NoStop}%
	\bibitem [{\citenamefont {Spohn}(2004)}]{spohn_2004}%
	\BibitemOpen
	\bibfield  {author} {\bibinfo {author} {\bibfnamefont {H.}~\bibnamefont
			{Spohn}},\ }\href {https://doi.org/10.1017/CBO9780511535178} {\emph {\bibinfo
			{title} {Dynamics of Charged Particles and their Radiation Field}}}\
	(\bibinfo  {publisher} {Cambridge University Press},\
	\bibinfo {address} {Cambridge, UK},\ \bibinfo {year}
	{2004})\BibitemShut {NoStop}%
	\bibitem [{\citenamefont {Schlosshauer}(2007)}]{Schlosshauer2007}%
	\BibitemOpen
	\bibfield  {author} {\bibinfo {author} {\bibfnamefont {M.~A.}\ \bibnamefont
			{Schlosshauer}},\ }\href {https://doi.org/10.1007/978-3-540-35775-9} {\emph
		{\bibinfo {title} {Decoherence and the Quantum-To-Classical Transition}}}\
	(\bibinfo  {publisher} {Springer-Verlag, Berlin},\ \bibinfo {year}
	{2007})\BibitemShut {NoStop}%
	\bibitem [{\citenamefont {Unruh}(2000)}]{Unruh_Coherence}%
	\BibitemOpen
	\bibfield  {author} {\bibinfo {author} {\bibfnamefont {W.~G.}\ \bibnamefont
			{Unruh}},\ }in\ \href
	{https://link.springer.com/chapter/10.1007/3-540-45369-5_7} {\emph {\bibinfo
			{booktitle} {Relativistic Quantum Measurement and Decoherence}}},\ \bibinfo
	{editor} {edited by\ \bibinfo {editor} {\bibfnamefont {H.-P.}\ \bibnamefont
			{Breuer}}\ and\ \bibinfo {editor} {\bibfnamefont {F.}~\bibnamefont
			{Petruccione}}}\ (\bibinfo  {publisher} {Springer},\
	\bibinfo {address} {Berlin},\ \bibinfo {year} {2000}),\ pp.\
	\bibinfo {pages} {125--140}\BibitemShut {NoStop}%
\end{thebibliography}
\begin{widetext}
\end{widetext}

\end{document}